\pdfoutput=1
\documentclass[useAMS,usenatbib]{mn2e}
\usepackage{apjfonts}
\usepackage{graphicx}
\usepackage{amsmath}
\usepackage{amssymb}
\usepackage{ctable}
\usepackage{fixltx2e}
\usepackage{threeparttable}
\usepackage{hyperref}
\hypersetup{colorlinks=true,linkcolor=blue,citecolor=blue,filecolor=blue,urlcolor=blue}

\newcommand{\be}{\begin{equation}}
\newcommand{\ee}{\end{equation}}
\newcommand{\ba}{\begin{eqnarray}}
\newcommand{\ea}{\end{eqnarray}}

\newcommand{\Mhalo}{M_{\rm halo}}

\newcommand{\Rhalf}{R_{\rm half}}
\newcommand{\Ms}{M_{\ast}}

\newcommand{\cm}{{\rm cm}}
\newcommand{\km}{{\rm km}}
\newcommand{\pc}{{\rm pc}}
\newcommand{\kpc}{{\rm kpc}}

\newcommand{\dex}{{\rm dex}}
\newcommand{\s}{{\rm s}}

\newcommand{\yr}{{\rm yr}}
\newcommand{\Myr}{{\rm Myr}}

\newcommand{\Msun}{M_{\sun}}

\newcommand{\Zsun}{Z_{\sun}}

\newcommand{\mb}{m_b}
\newcommand{\mDM}{m_{\rm DM}}
\newcommand{\eg}{\epsilon_{\rm gas}}
\newcommand{\es}{\epsilon_{\rm star}}
\newcommand{\eDM}{\epsilon_{\rm DM}}

\newcommand{\nc}{n_{\rm th}}

\newcommand{\Mc}{M_{\rm cl}}

\newcommand{\dd}{{\rm d}}
\newcommand{\LCDM}{$\Lambda$CDM}
\newcommand{\mol}{{\sc mol}}
\newcommand{\sg}{{\sc sg}}
\newcommand{\den}{{\sc den}}
\newcommand{\cf}{{\sc cf}}
\newcommand{\eff}{\epsilon_{\rm ff}}
\newcommand{\aj}{AJ}
\newcommand{\apj}{ApJ}
\newcommand{\apjl}{ApJ}
\newcommand{\apjs}{ApJS}
\newcommand{\mnras}{MNRAS}
\newcommand{\aap}{A\&A}
\newcommand{\araa}{ARA\&A}

\newcommand{\rmxaa}{RMxAA}
\newcommand{\physrep}{Physics Reports}

\title[GCs in high-$z$ galaxies]
{Self-consistent proto-globular cluster formation in cosmological simulations of high-redshift galaxies}

\author[X. Ma et al.]{
  \parbox[t]{1.0\textwidth}{
   Xiangcheng Ma,$^{1}$\thanks{E-mail: xchma@berkeley.edu}
   Michael Y. Grudi{\'c},$^2$
   Eliot Quataert,$^1$
   Philip F. Hopkins,$^2$ \\
   Claude-Andr{\'e} Faucher-Gigu{\`e}re,$^3$
   Michael Boylan-Kolchin,$^4$
   Andrew Wetzel,$^5$ \\
   Ji-hoon Kim,$^6$
   Norman Murray$^7$ and
   Du{\v s}an Kere{\v s}$^8$
  }
  \vspace{5pt} \\
  $^1$Department of Astronomy and Theoretical Astrophysics Center, University of California Berkeley, Berkeley, CA 94720 \\
  $^2$TAPIR, MC 350-17, California Institute of Technology, Pasadena, CA 91125, USA \\ 
  $^3$Department of Physics and Astronomy and CIERA, Northwestern University, 2145 Sheridan Road, Evanston, IL 60208, USA \\
  $^4$Department of Astronomy, The University of Texas at Austin, 2515 Speedway Blvd, Stop C1400, Austin, TX 78712, USA \\
  $^5$Department of Physics, University of California, Davis, CA 95616, USA \\
  $^6$Center for Theoretical Physics, Department of Physics and Astronomy, Seoul National University, Seoul 08826, Korea \\
  $^7$Canadian Institute for Theoretical Astrophysics, 60 St George Street, University of Toronto, ON M5S 3H8, Canada \\
  $^8$Department of Physics, Center for Astrophysics and Space Sciences, University of California at San Diego, 9500 Gilman Drive, La Jolla, CA 92093
}

\pagerange{\pageref{firstpage}--\pageref{lastpage}}
\date{Draft version \today}

\begin{document}
\maketitle
\label{firstpage}

\begin{abstract}
We report the formation of bound star clusters in a sample of high-resolution cosmological zoom-in simulations of $z\geq5$ galaxies from the FIRE project. We find that bound clusters preferentially form in high-pressure clouds with gas surface densities over $10^4\,\Msun\,\pc^{-2}$, where the cloud-scale star formation efficiency is near unity and young stars born in these regions are gravitationally bound at birth. These high-pressure clouds are compressed by feedback-driven winds and/or collisions of smaller clouds/gas streams in highly gas-rich, turbulent environments. The newly formed clusters follow a power-law mass function of $\dd N/\dd M \sim M^{-2}$. The cluster formation efficiency is similar across galaxies with stellar masses of $\sim10^7$--$10^{10}\,\Msun$ at $z\geq5$. The age spread of cluster stars is typically a few Myrs and increases with cluster mass. The metallicity dispersion of cluster members is $\sim0.08$\,dex in $\rm [Z/H]$ and does not depend on cluster mass significantly. Our findings support the scenario that present-day old globular clusters (GCs) were formed during relatively normal star formation in high-redshift galaxies. Simulations with a stricter/looser star formation model form a factor of a few more/fewer bound clusters per stellar mass formed, while the shape of the mass function is unchanged. Simulations with a lower {\em local} star formation efficiency form more stars in bound clusters. The simulated clusters are larger than observed GCs due to finite resolution. Our simulations are among the first cosmological simulations that form bound clusters self-consistently in a wide range of high-redshift galaxies.
\end{abstract}


\section{Introduction}
\label{sec:intro}
Globular clusters (GCs) are spherical, tightly bound collections of stars that are typically 5--13\,Gyr old, massive ($10^4$--$10^6\,\Msun$), metal-poor ($\rm -2.5 \lesssim [Fe/H] \lesssim 0.3$), and compact (with half-light radii of 0.5--10\,pc) \citep[see e.g.][for a series of reviews]{Harris:1991,Brodie:2006,Bastian:2018}. GCs are found in almost all galaxies above $\Ms\sim10^9\,\Msun$ in the local Universe.

The origin of GCs remains a longstanding puzzle in the context of star formation and galaxy formation. A number of theories have been proposed to explain their formation. For example, \citet{Peebles:1968} suggested that GCs were the first bound structures formed in the universe when the Jean mass following recombination is $10^5$--$10^6\,\Msun$. Similarly, \citet{Fall:1985} suggested that GCs were formed via thermal instabilities of metal-poor gas in galactic halos where the Jeans mass is $10^6\,\Msun$ \citep[see also][]{Kang:1990,Shapiro:1992}. \citet{Peebles:1984} has suggested that GCs were formed in dark matter (DM) halos at high redshift that merged into more massive halos at later times. Recent cosmological simulations have shown that GC-like objects can form in $10^6$--$10^8\,\Msun$ halos in the very early universe \citep[e.g.][]{Trenti:2015,Kimm:2016,Ricotti:2016}. \citet{Naoz:2014} suggested GCs can form without a DM halo in the early Universe due to the relative stream velocity between DM and baryons induced at the epoch of recombination \citep[see also e.g.][]{Chiou:2019}. Some GCs may also be the nuclei or nuclear clusters of former dwarf galaxies that were tidally disrupted \citep[e.g.][]{Mackey:2005}.

Since the 1980s, a population of stellar systems called young massive clusters (YMCs) have been found in some extreme environments in the local Universe, including galaxy mergers, starburst galaxies, and galactic nuclei \citep[see][for a recent review]{Portegies-Zwart:2010}. The YMCs are younger than 1\,Gyr, with masses of $10^4$--$10^8\,\Msun$ and half-light radii only a few pc \citep[e.g.][]{Larsen:2004}. The YMCs broadly follow a mass function of $\dd N/\dd M\sim M^{-2}$ with a truncation at the high-mass end \citep[e.g.][]{Zhang:1999,Gieles:2006,Larsen:2009}. Given the similarities between GCs and YMCs, many authors have suggested that present-day GCs were in fact YMCs formed at early times \citep[see][for recent reviews]{Longmore:2014,Kruijssen:2014}.

YMCs are preferentially formed in highly gas-rich, turbulent conditions, with gas surface densities of $\Sigma_{\rm gas}\sim10^2$--$10^4\,\Msun\,\pc^{-2}$ and turbulent velocities of $\sigma_v\sim10$--$100\,\km\,\s^{-1}$, translating to pressures of $P/k\sim10^6$--$10^8$\,K\,$\cm^{-3}$ \citep[see e.g.][]{Longmore:2014}.\footnote{Throughout this paper, when we refer to a `high-pressure' region where bound clusters form, by `pressure' we mean the gravitational force per unit area $P_{\rm grav}\sim G\Sigma_{\rm gas}^2$. This can be balanced by turbulent ram pressure $P_{\rm turb} \sim \rho\sigma_v^2$, which is comparable to $P_{\rm grav}$ \citep[following][]{Elmegreen:1997}. We do not mean thermal pressure as the gas is usually cold.} Such pressures are almost three orders of magnitude higher than those in our Milky Way (MW) and typical disk galaxies in the local Universe \citep[see e.g.][]{Kruijssen:2013a}, but are commonly seen in high-redshift galaxies \citep[e.g.][and references therein]{Kruijssen:2014}. Both theories and numerical simulations showed that in such high-pressure regions, the integrated star formation efficiency over a cloud (i.e. the fraction of the cloud mass that turns into stars) can reach above 50\% and hence a large fraction of stars from in bound clusters \citep[e.g.][for recent studies]{Fall:2010,Skinner:2015,Kim:2016a,Grudic:2018,Tsang:2018,Li:2019}.\footnote{For example, \cite{Grudic:2018} found the integrated star formation efficiency of a cloud is set by its surface density, $\epsilon_{\rm int}\sim(1+\Sigma_{\rm crit}/\Sigma_{\rm gas})^{-1}$. This can be derived assuming star formation in the cloud stops when feedback from newly formed stars becomes larger than the self-gravity of the cloud. They found $\Sigma_{\rm crit}\sim3000\,\Msun\,\pc^{-2}$ using high-resolution cloud-scale simulations. Most stars form in one free-fall time of the cloud, so $\epsilon_{\rm int}$ is also the star formation efficiency per {\em cloud} free-fall time. We will refer back to these results later in this paper.} This further supports the idea that GCs form in regular star formation in high-redshift galaxies where bound clusters can form efficiently \citep[see e.g.][]{Elmegreen:1997,Elmegreen:2012,Kruijssen:2012b}.

Resolving the formation of YMCs or proto-GCs in cosmological simulations of galaxy formation is a challenging problem. This is due to the multi-scale nature of this task: a wide dynamic range is required to simultaneously capture cosmic structure formation, hierarchical fragmentation of the interstellar medium (ISM), and star formation and feedback physics in giant molecular clouds (GMCs). State-of-the-art cosmological simulations specifically designed to study GC formation and evolution down to $z=0$ cannot yet resolve substructures in the GMCs. A commonly adopted approach is to spawn tracer particles to represent bound clusters. When and where cluster particles form are determined empirically according to local gas properties (see e.g. \citealt{Kravtsov:2005,Li:2017}; the E-MOSAICS simulations, \citealt{Pfeffer:2018,Kruijssen:2019}). Such models may include free parameters for physics on under-resolved scales \citep[e.g.][]{Li:2018}.

A different family of models for GC formation and evolution across cosmic time use semi-analytic approach built upon DM halo merger trees \citep[e.g.][]{Beasley:2002,Muratov:2010,Li:2014,Kruijssen:2015,Choksi:2018,El-Badry:2019}. These models usually determine GC formation rates by halo properties and/or galaxy-scale gas conditions, such as merger rates \citep[e.g.][]{Li:2014}, halo growth rates \citep[e.g.][]{Choksi:2018}, and average gas surface densities \citep[e.g.][]{El-Badry:2019}. These models are complementary to cosmological simulations.

Thanks to the dramatically increased speed of supercomputers over the past few years, cosmological simulations now can achieve unprecedented resolution. In the meanwhile, models used in state-of-the-art simulations have also been substantially improved, making it possible to better capture the multi-phase ISM, star formation in GMCs, disruption of GMCs by stellar feedback, and the launch and propagation of galactic winds, at least in high-resolution cosmological zoom-in simulations \citep[e.g.][]{Ceverino:2014,Hopkins:2014,Hopkins:2018b,Wang:2015,Agertz:2016}. Recently, \citet{Kim:2018} reported a proto-GC formed during a gas-rich merger of two low-mass $z\sim7$ galaxies in a cosmological zoom-in simulation from the Feedback in Realistic Environments (FIRE) project.\footnote{\url{https://fire.northwestern.edu}} \citet{Mandelker:2018} found that GC-like objects formed via fragmentations of cold streams from the cosmic web in a cosmological zoom-in simulation of a $z\sim6$ galaxy. 

In this paper, we study a suite of high-resolution cosmological zoom-in simulations of galaxies at $z\geq5$, where we use physically motivated models for the multi-phase ISM, star formation, and stellar feedback. The simulations are only run to $z=5$, so that we can use sufficiently high mass resolution at a reasonable computational cost. The goal of this paper is to show that YMCs/proto-GCs form in our simulations over a wide range of galaxy masses and identify the key processes resulting in cluster formation. We will show that newly formed clusters broadly follow a power-law mass function of $\dd N/\dd M\sim M^{-2}$ as arises naturally from hierarchical structure formation \citep[e.g.][]{Elmegreen:1996,Fujii:2013,Guszejnov:2018}. Our simulations are among the first to form bound clusters in `regular' star formation in gas-rich high-redshift galaxies self-consistently. 

Being able to explicitly resolve the formation of YMCs/proto-GCs in cosmological simulations is important for the following reasons. First of all, it complements high-resolution cloud-scale simulations of cluster formation \citep[e.g.][and references above]{Mac-Low:2004,McKee:2007,Krumholz:2014} by bridging small- and large-scale simulations together. Secondly, it is also complementary to previous studies on cluster formation in analytic models \citep[e.g.][]{Elmegreen:1997,Kruijssen:2012b,Grudic:2018a} and idealized simulations of isolated galaxies and gas-rich mergers \citep[e.g.][]{Li:2004,Kruijssen:2011,Kruijssen:2012,Hopkins:2012,Hopkins:2013,Maji:2017,Lahen:2019,Moreno:2019} by adding a cosmological environment. Finally, doing so will provide insights and calibrate empirical models implemented in other simulations and semi-analytic models.

\begin{table*}
\caption{Initial conditions of simulations studied in this paper.}
\begin{threeparttable}
\begin{tabular}{lcccccccc}
\hline
Simulation & $\Mhalo$ ($z=5$) & $\Ms$ ($z=5$) & $\mb$ & $\eg$ & $\es$ & $\mDM$ & $\eDM$ & Notes \\
 & [$\Msun$] & [$\Msun$] & [$\Msun$] & [pc] & [pc] & [$\Msun$] & [pc] & \\
\hline
z5m10a\_hr & $6.6\times10^9$ & $1.5\times10^7$ & 119.3 & 0.14 & 0.7 & 6.5e2 & 10 \\
z5m11c\_hr & $7.6\times10^{10}$ & $9.4\times10^8$ & 890.8 & 0.28 & 1.4 & 4.9e3 & 21 & 8 times better mass resolution than z5m11c \\
z5m11c & $7.8\times10^{10}$ & $7.8\times10^8$ & 7126.5 & 0.42 & 2.1 & 3.9e4 & 42 & 8 times lower mass resolution than z5m11c\_hr \\
z5m12b & $8.6\times10^{11}$ & $2.2\times10^{10}$ & 7126.5 & 0.42 & 2.1 & 3.9e4 & 42 \\
\hline
\end{tabular}
\begin{tablenotes}
\item Parameters describing our simulations:
\item (1) $\Mhalo$: Halo mass of the central halo in the zoom-in region.
\item (2) $\Ms$: Total stellar mass inside the virial radius of the central halo.
\item (3) $\mb$ and $\mDM$: Initial baryonic and DM particle masses in the high-resolution region. 
\item (4) $\eg$, $\es$, and $\eDM$: Plummer-equivalent force softening lengths for gas, star, and DM particles, in comoving units above $z=9$ and physical units thereafter. Force softening for gas is adaptive ($\eg$ is the minimum softening length).
\end{tablenotes}
\end{threeparttable}
\label{tbl:zoom}
\end{table*} 

Understanding cluster formation in $z\geq5$ galaxies is also important for understanding the galaxy populations at these redshifts and the reionization history. It has been suggested that the progenitors of present-day GCs may have contributed significantly to the cosmic star formation rate densities at $z\geq5$ and thus the ionizing photon budgets available for cosmic reionization \citep[e.g.][]{Boylan-Kolchin:2018,Zick:2018}. Recently, \citet{Bouwens:2017b} reported that dwarf galaxies at $z\sim6$--8 found in the Hubble Frontier Fields (HFFs) have very small sizes, with individual sources reaching as small as $\sim10$--30\,pc, similar to star cluster complexes and/or super star clusters seen in the local Universe \citep[see also e.g.][]{Bouwens:2017,Kawamata:2018}. \citet{Vanzella:2017,Vanzella:2017a,Vanzella:2019} also reported a superdense star-forming region at $z=6.143$ with an effective radius less than 13\,pc and a star formation rate density of $\sim1000\,\Msun\,\yr^{-1}\,\kpc^{-2}$ behind one of the HFF clusters. It is quite possible that these small objects in the early Universe are newly formed star clusters/YMCs/proto-GCs that may eventually become present-day GCs. These clusters might bias the interpretation of the faint-end galaxy ultraviolet luminosity functions  (UVLFs) at $z\geq5$, as they can be preferentially selected in surface-brightness-limited surveys \citep[see e.g.][]{Bouwens:2017b,Ma:2018}. 

This paper is organized as follows. In Section \ref{sec:method}, we introduce our simulation sample and describe the star formation and feedback physics adopted in these simulations. In Section \ref{sec:formation}, we present case studies to show the key processes involved in cluster formation. We also present the mass functions for newly formed clusters in these simulations. In Section \ref{sec:property}, we show the properties of these clusters. In Section \ref{sec:sf}, we show how our results are affected by the details in our star formation model. We discuss the implications of our results in Section \ref{sec:discussion} and conclude in Section \ref{sec:conclusion}.

We adopt a standard flat {\LCDM} cosmology with {\it Planck} 2015 cosmological parameters $H_0=68 {\rm\,km\,s^{-1}\,Mpc^{-1}}$, $\Omega_{\Lambda}=0.69$, $\Omega_{m}=1-\Omega_{\Lambda}=0.31$, $\Omega_b=0.048$, $\sigma_8=0.82$, and $n=0.97$ \citep{Planck-Collaboration:2016a}. We use a \citet{Kroupa:2002} initial mass function (IMF) from 0.1--$100\,\Msun$, with IMF slopes of $-1.30$ from 0.1--$0.5\,\Msun$ and $-2.35$ from 0.5--$100\,\Msun$.

\begin{table*}
\caption{Simulation restarts. From each parent simulation, a starburst is re-started and run for the cosmic time and redshift labeled (see Fig. \ref{fig:sfr}) with different star formation models. We consider the following criteria for star formation: (i) molecular (\mol), (ii) self-gravitating (\sg), (iii) density threshold (\den), and (iv) converging flow (\cf) (as defined in Sections \ref{sec:zoom} and \ref{sec:restart}). Each model name refers to a combination of star formation criteria with certain choice of local star formation efficiency $\eff$ listed below (see Section \ref{sec:restart} for details). We refer to these re-simulations by {\em Parent simulation\_Model name} (e.g. z5m11c\_hr\_G18). More detailed comparison and discussion between different star formation models are presented in Section \ref{sec:sf}.}
\begin{threeparttable}
\begin{tabular}{lcccccccc}
\hline
Parent & Redshift & Cosmic time & $\Mhalo$ & Model & Star formation & $\eff$ & $\Ms$ formed & $f_{\rm bound}$ \\
simulation & & [Gyr] & [$\Msun$] & name & criteria & & [$\Msun$] & \\
\hline
z5m10a\_hr & 6.605--6.143 & 0.820--0.901 & $5.8\times10^9$ & G18 & {\mol+\sg+\cf} & 1 & $1.57\times10^7$ & 0.26 \\
z5m11c\_hr & 5.562--5.024 & 1.023--1.163 & $7.6\times10^{10}$ & FIRE & {\mol+\sg+\den} & 1 & $2.56\times10^8$ & 0.28  \\
 & & & & no $\nc$ & {\mol+\sg} & 1 & $3.50\times10^8$ & 0.17 \\
 & & & & G18 & {\mol+\sg+\cf} & 1 & $4.44\times10^8$ & 0.26 \\
 & & & & G18\_e50 & {\mol+\sg+\cf} & 0.5 & $3.00\times10^8$ & 0.39 \\
z5m11c & 5.562--5.024 & 1.023--1.163 & $7.6\times10^{10}$ & G18 & {\mol+\sg+\cf} & 1 & $1.55\times10^8$ & 0.12 \\
z5m12b & 6.333--6.080 & 0.866--0.913 & $4.5\times10^{11}$ & G18 & {\mol+\sg+\cf} & 1 & $3.31\times10^9$ & 0.33 \\
\hline
\end{tabular}
\begin{tablenotes}
\item {\em Notes.} (1) Parent simulation: The zoom-in simulation where the starburst is selected (see Table \ref{tbl:zoom} and Fig. \ref{fig:sfr}). (2) Redshift and cosmic time: The duration in redshift and cosmic time where the simulation is re-run. (3) $\Mhalo$: Average halo mass during the re-simulation. (4) $\Ms$ formed: Total stellar mass formed in the galaxy during the restart. (5) $f_{\rm bound}$: The fraction of stars formed during the re-simulation that end up in bound clusters by the end of the re-simulation.
\end{tablenotes}
\end{threeparttable}
\label{tbl:restart}
\end{table*}

\begin{figure*}
\centering
\includegraphics[width=0.70\linewidth]{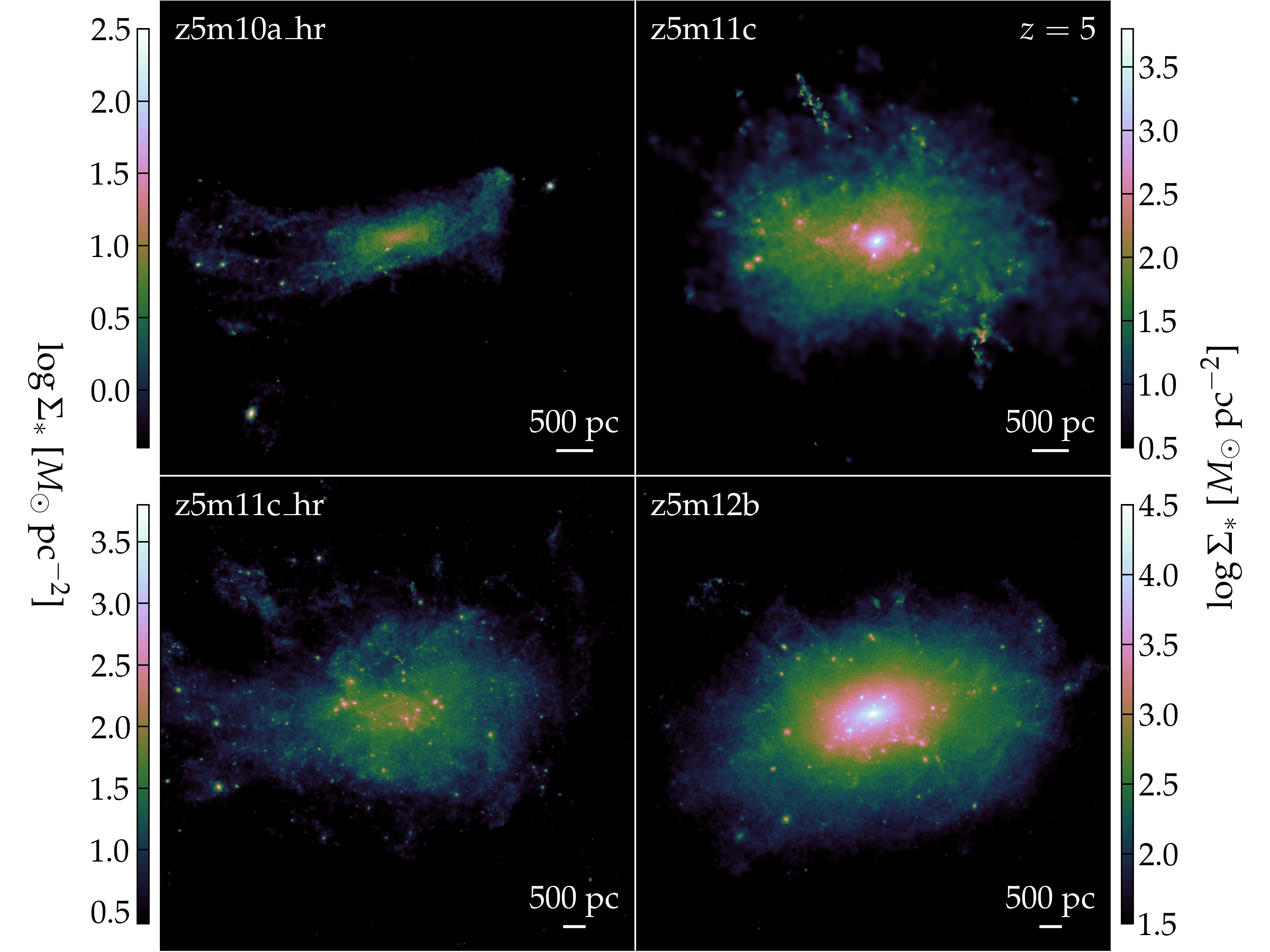}
\caption{Stellar surface density maps of the four galaxies from the FIRE-2 simulation suite (Table \ref{tbl:zoom}), viewed at the final redshift of the simulations ($z=5$). All galaxies contain a number of small, GC-like objects at least survived by $z=5$.}
\label{fig:image} \vspace{3pt}

\includegraphics[width=\linewidth]{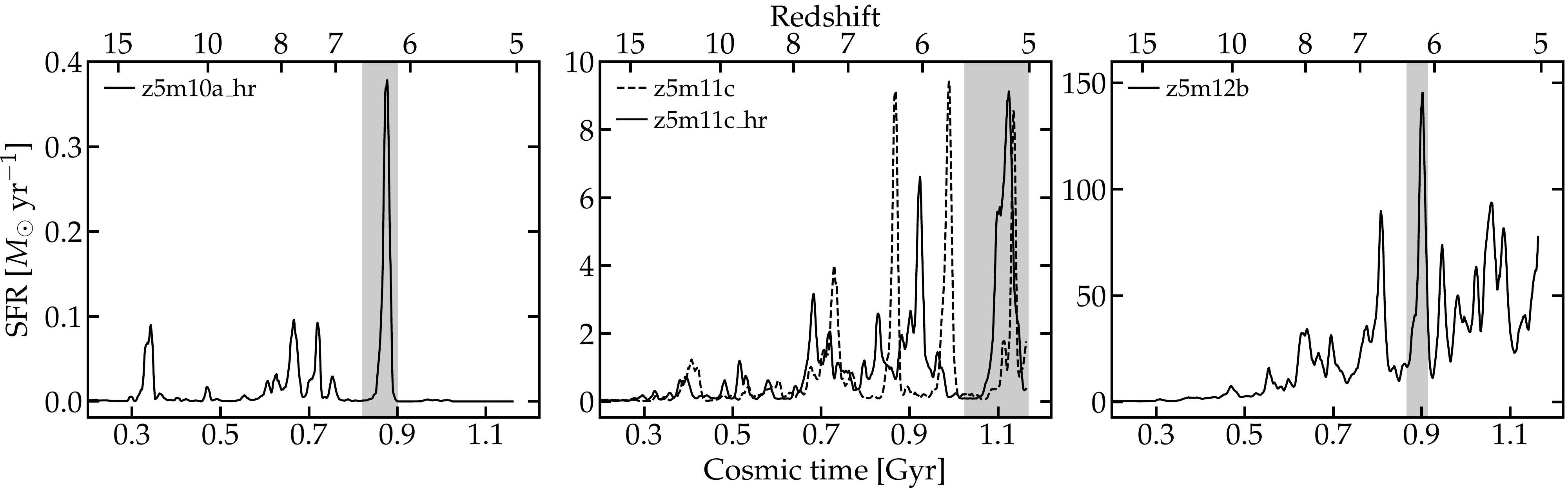}
\caption{Star formation histories of the four simulated galaxies in Fig. \ref{fig:image}. All galaxies undergo bursty star formation on time-scales of $\sim100\,\Myr$. Most stars (and hence clusters) in the galaxies are formed during these starbursts. For each galaxy, we re-simulate one starburst with much finer output (0.5\,Myr between consecutive snapshots) to study the formation of GC candidates. The grey shaded regions illustrate the duration of the restart for each galaxy. We also consider various star formation models in the re-simulations and compare the results in Section \ref{sec:sf}.}
\label{fig:sfr} 
\end{figure*}

\section{Methods}
\label{sec:method}

\subsection{Cosmological zoom-in simulations}
\label{sec:zoom}
Our study is based on a sample of cosmological zoom-in simulations from the FIRE project, which we summarize in Table \ref{tbl:zoom}. The initial conditions for halos z5m10a, z5m11c, and z5m12b are first introduced in \cite{Ma:2018a}. These three halos are selected at $z=5$ with approximate halo mass $10^{10}$, $10^{11}$, and $10^{12}\,\Msun$, respectively, at this redshift. We have run halos z5m10a and z5m11c with 8 times better mass resolution than those used in our previous work, which we refer to as z5m10a\_hr and z5m11c\_hr. We include halo z5m11c run at two different resolution for comparison. Halos z5m11c and z5m12b are identical to those used in \citet{Ma:2019}. In Table \ref{tbl:zoom}, we list the mass resolution and force softening lengths for baryonic and dark matter (DM) particles adopted in these simulations. All the multi-scale cosmological zoom-in initial conditions are generated with the code {\sc music} \citep{Hahn:2011}.

The simulations are run using an identical version of the code {\sc gizmo}\footnote{\url{http://www.tapir.caltech.edu/~phopkins/Site/GIZMO.html}} \citep{Hopkins:2015} in the mesh-less finite-mass (MFM) mode. We use the FIRE-2 models for the multi-phase ISM, star formation, and stellar feedback. We briefly review and highlight the key ingredients that are important to this study below, but refer to \cite{Hopkins:2018a,Hopkins:2018b} for more details of the numerical implementation and tests. Gas follows an ionized+atomic+molecular cooling curve between 10--$10^{10}$\,K, including metallicity-dependent fine-structure and molecular cooling at low temperatures and metal-line cooling at high temperatures. The ionization states and cooling rates for H and He are computed following \cite{Katz:1996} and cooling rates for heavy elements are calculated from a compilation of {\sc cloudy} runs \citep{Ferland:2013}, applying a uniform, redshift-dependent ionizing background from \cite{Faucher-Giguere:2009} and heating from local sources. Self-shielding is accounted for with a local Sobolev/Jeans-length approximation.

Star formation is allowed only if the following criteria are met: 

(i) {\em Molecular} (\mol). The self-shielded molecular fraction of each gas particle ($f_{\rm mol}$) is estimated following \cite{Krumholz:2011}. Star formation only takes place in molecular gas.

(ii) {\em Self-gravitating} (\sg). This requires that the gravitational potential energy dominates over kinetic plus thermal energy at the resolution scale, as described by the virial parameter
\be
\label{eqn:virial}
\alpha \equiv \frac{\| \nabla\otimes\mathbf{v} \|^2_i + (c_{{\rm s},\,i}/h_i)^2}{8\pi G \, \rho_i} < 1,
\ee
where $\otimes$ is the outer product, $c_{\rm s}$ is the sound speed, $h$ is the resolution scale, and the subscript $i$ implies that the quantities are evaluated for individual gas particles \citep{Hopkins:2013b}.

(iii) {\em Density threshold} (\den). The number density of hydrogen exceeds a threshold of $n_{\rm H}>\nc=1000\,\cm^{-3}$. 

If all criteria above are met, a gas particle will turn into a star particle at a rate $\dot{\rho}_{\ast}=\eff \, f_{\rm mol} \,\rho/t_{\rm ff}$, where $\eff$ is the {\em local} star formation efficiency and $t_{\rm ff}$ is the free-fall time of the {\em particle}. We adopt $\eff=1$ by default. We note that this efficiency reflects the rate at which locally self-gravitating clumps fragment, while the realized star formation efficiency of GMCs is regulated by feedback at $\sim1$--10\% per {\em cloud} free-fall time for typical MW GMC conditions \citep[see e.g.][and reference therein]{Hopkins:2018b}. Only in extremely high-surface-density regions ($\gg10^3\,\Msun\,\pc^{-2}$) where feedback is no longer sufficiently strong to overcome gravity can the cloud-scale star formation efficiency reach near unity \citep[e.g.][]{Grudic:2018}. We will vary $\eff$ and consider criteria for star formation different from the FIRE-2 model in Section \ref{sec:restart}.

Every star particle is treated as a single stellar population with known age, mass, and metallicity (inherited from its parent gas particle). All feedback quantities are calculated directly from standard stellar population synthesis models {\sc starburst99} \citep{Leitherer:1999} assuming a \cite{Kroupa:2002} IMF. The simulations account for the following feedback mechanisms: (1) photoionization and photoelectric heating, (2) radiation pressure for UV/optical single scattering and multiple scattering of re-radiated IR photons (the latter is usually subdominant in the simulations), and (3) energy, momentum, mass, and metal injection from discrete supernovae and continuous stellar winds (from both OB and AGB stars). More details on the numerics of the radiative feedback (1 and 2) and mechanical feedback (3) are presented in \cite{Hopkins:2018c,Hopkins:2018a}. We include metal yields from Type-II SNe, Type-Ia SNe, and AGB winds \citep[see][Appendix A]{Hopkins:2018b}.

The simulations adopt the sub-resolution turbulent metal diffusion and mixing model described in \cite{Su:2017} and \cite{Escala:2018}. We do not model the chemistry appropriate for primordial gas nor the formation and evolution of Pop\,{\sc iii} stars, but assume a metallicity floor at $Z=10^{-4}\,\Zsun$.

\subsection{Simulation restarts}
\label{sec:restart}
Fig. \ref{fig:image} shows the stellar surface density maps for the four simulated galaxies in Table \ref{tbl:zoom} at the final redshift of these simulations ($z=5$). Every galaxy contains a number of small, cluster-like objects which survive at least to $z=5$. The central surface density of the clusters in galaxy z5m10a\_hr is $\sim100$--$300\,\Msun\,\pc^{-2}$, while those in galaxy z5m12b typically have central surface densities over $10^4\,\Msun\,\pc^{-2}$. Galaxy z5m11c\_hr shows more smaller clusters than z5m11c owing to its higher mass resolution. Fig. \ref{fig:sfr} shows the star formation history of the four galaxies. All galaxies exhibit bursty star formation to some extent on time-scales of $\sim100\,\Myr$ \citep[see also][fig. 5]{Ma:2018a}. We note that when and where a gas particle turns into a star particle and a SN occurs are sampled stochastically from the SFR and SN rates in our simulations, so running the same simulation twice does {\em not} lead to identical results \citep{Hopkins:2018b}. The difference between z5m11c and z5m11c\_hr in their morphologies and star formation histories is likely due to stochastic effects. They should be regarded as two realizations of the same halo.

Most of the stars and hence clusters in the galaxies are formed during these bursts of star formation. The time-scale of cluster formation is of the same order as the free-fall time of its parent cloud, which is at most a few Myrs for typical densities of GCs \citep[e.g.][]{Kim:2018,Grudic:2018}. The default time interval between consecutive snapshots in our simulations ($\sim20\,\Myr$) is too large for studying the formation of GCs, so for each simulation, we restart the run from a snapshot prior to a starburst with more frequent outputs ($\sim0.5$\,Myr between snapshots) until the end of the burst. The grey shaded regions in Fig. \ref{fig:sfr} illustrate the starbursts we re-simulate for each galaxy. The starting and final cosmic time and redshift for each burst are listed in Table \ref{tbl:restart}. Galaxies z5m11c and z5m11c\_hr both exhibit a starburst slightly before $z=5$, so we re-run them for the same redshift interval. We point out that the starburst we select for z5m12a\_hr is triggered by a major merger, while those in other simulations are not associated with mergers.

We consider alternative star formation models and parameters when re-running the starbursts to understand the robustness of cluster formation in our simulations. In addition to the star formation criteria (i)--(iii) above, we also consider a fourth criterion, requiring a locally convergent flow:

(iv) {\em Converging flow} (\cf). Star formation is only allowed in converging flows where $\nabla\cdot{\bf v}<0$.

Each star formation model combines some of the above criteria and certain choice of $\eff$. In Table \ref{tbl:restart}, we list the star formation models we explore in this paper. The standard FIRE-2 star formation model consists of criteria {\mol+\sg+\den} with local star formation efficiency $\eff=1$. The `no $\nc$' model only includes criteria {\mol+\sg}, i.e. no density threshold for star formation compared to the FIRE-2 model. The `G18' model is adopted in \citet{Grudic:2018}, which consists of criteria {\mol+\sg+\cf}. The `G18\_e50' model is identical to `G18' model, except that we adopt a lower local star formation efficiency of $\eff=0.5$. In Sections \ref{sec:formation} and \ref{sec:property}, we choose model `G18' as our default model when studying GC formation during these starbursts. Grudi{\'c} et al. (2019) showed that this model reproduces the observed fraction of stars forming in bound clusters in a suite of high-resolution simulations of individual clouds similar to MW and M51 GMCs. This model also gives good convergence with mass resolution. In Section \ref{sec:sf}, we will study how different star formation models affect our results.\footnote{The virial parameter $\alpha$ defined in Equation \ref{eqn:virial} usually exhibits some time variability due to random fluctuations in local gas motion. To avoid spurious star formation, \citet{Grudic:2018} also required $\alpha$ to change smoothly with time. This makes a small difference in the cluster mass function shown in Section \ref{fig:mf}, within stochastic effects (less than a factor of 2). We find that the converging-flow criterion plays a more dominant and systematic effect.}

\begin{figure*}
\centering
\includegraphics[width=0.85\linewidth]{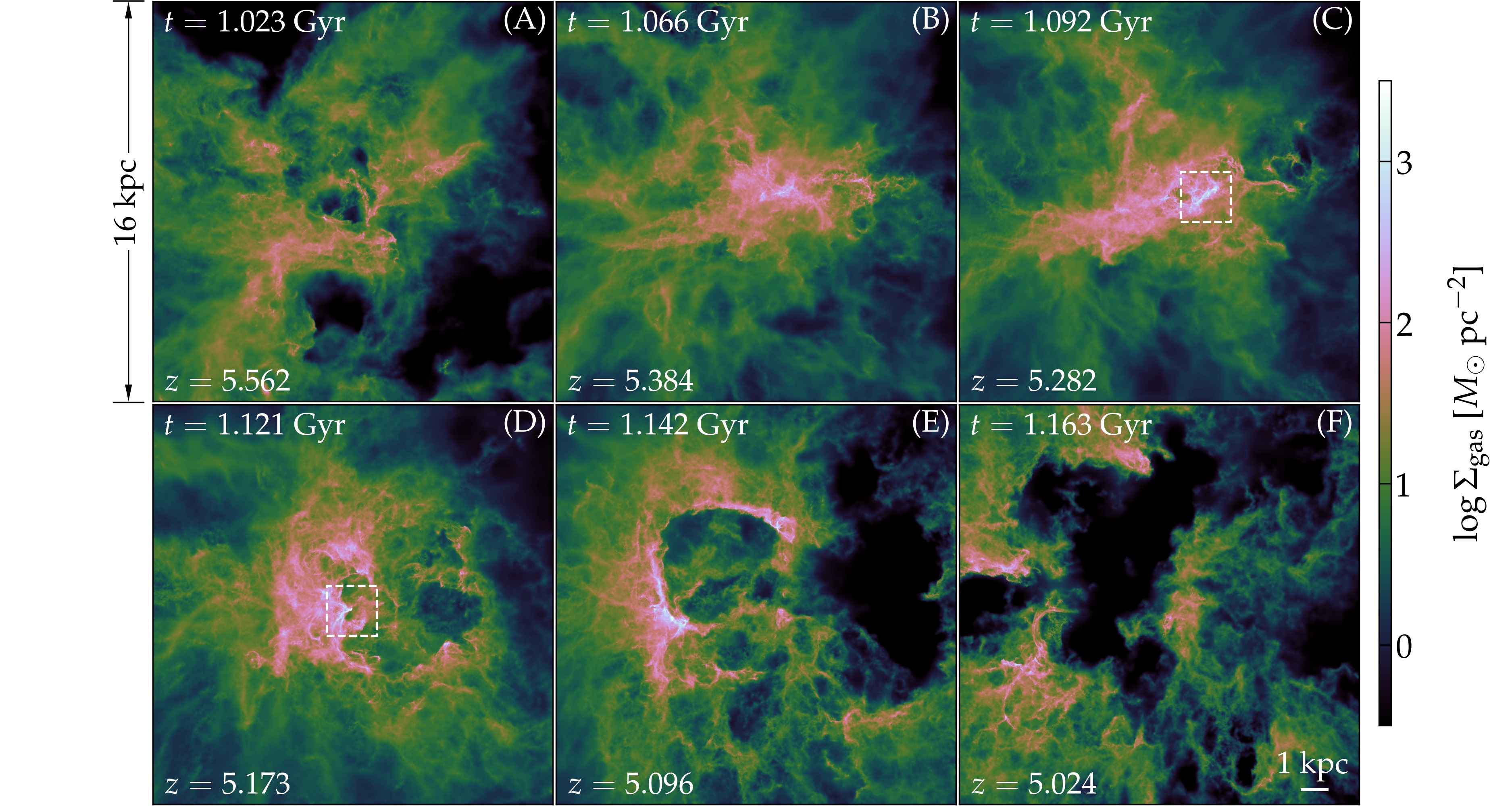}
\caption{Galactic-scale gas morphologies at six epochs (as labeled by the cosmic time and redshift) during the starburst in simulation z5m11c\_hr\_G18. Each panel is 16\,kpc along each side. (A) Gas flows in rapidly prior to the burst. (B) A large amount of gas has built up in the central region. (C) Star formation is triggered in dense clouds. (D--E) Feedback from recently formed stars drives gas flows in the galaxy. The gas flows compress nearby gas to very high densities as they sweep across the ISM. (F) Strong galactic winds finally terminate the burst. The galaxy maintains an extremely high gas fraction (nearly 80\% within the central 5\,kpc) during the starburst. The gas is highly turbulent with typical Mach number ${\cal M}\sim10$--30. The white dashed boxes in panels C and D show a $2\,\kpc\times2\,\kpc$ region centered on the clusters in Figs. \ref{fig:c13} and \ref{fig:c00}, respectively, at their formation time.}
\label{fig:galaxy} 
\end{figure*}

\begin{figure*}
\includegraphics[width=0.83\linewidth]{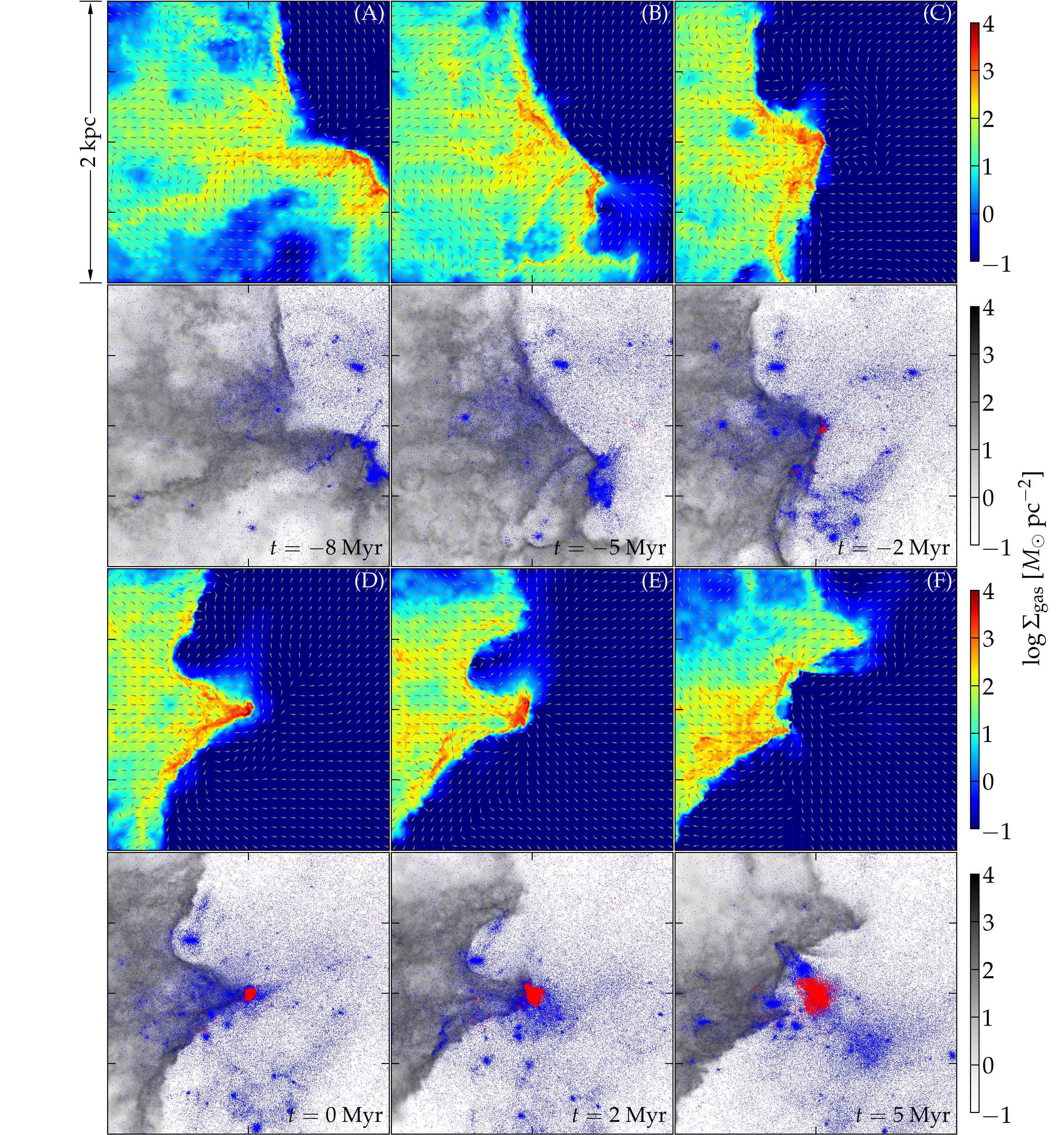}
\caption{The formation process of a proto-GC formed in z5m11c\_hr\_G18 (see the box in panel D of Fig. \ref{fig:galaxy}), which has a final mass of $\Mc=1.4\times10^7\,\Msun$ by the end of the simulation. Each panel is 2\,kpc along each side showing the projected image of a 500\,pc-thick slab of the ISM, centered on the median position of all cluster members (and/or their progenitor gas particles). The color (grey) maps in the first/third (second/fourth) rows show the gas surface density. The vector fields in the first/third rows ate the gas motion relative to the center of mass of cluster members (and/or their progenitors). The blue points in the second/fourth rows show all star particles in the slab, while the red points highlight the cluster members. The formation time is when half of the cluster members have formed ($t=0$). Panels A and B show a gas flow sweeping across the slab from top right to the bottom left. The dense gas entrained in the flow will collide with the dense gas along its path, forming a high-surface-density cloud at the center of panel C. Most cluster stars form (panels C--E) when the cloud maintains a surface density above $10^4\,\Msun\,\pc^{-2}$. This is sufficiently high that feedback fails to overcome the self-gravity of the cloud and almost all the gas in the cloud turns into stars within a cloud free-fall time. In the meanwhile, the cloud is fed by a large-scale flow from the left side of the frame, so the cloud can survive for a few Myrs. Feedback from the cluster finally blows out the surrounding gas in panel F.}
\label{fig:c00}
\end{figure*}

\begin{figure*}
\includegraphics[width=0.83\linewidth]{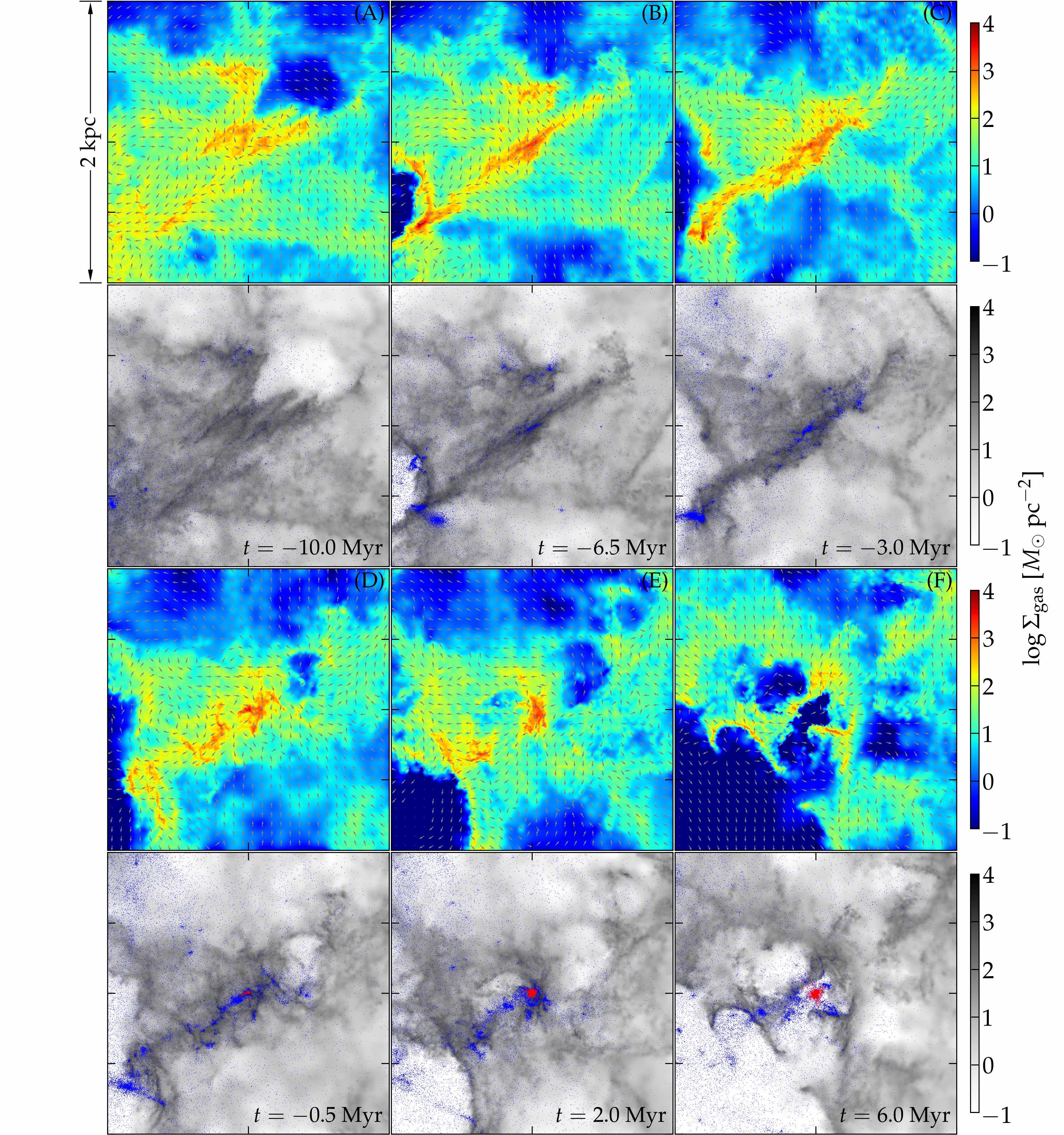}
\caption{The formation of another proto-GC in z5m11c\_hr\_G18 that has a final mass of $\Mc=1.6\times10^6\,\Msun$ (see also panel C of Fig. \ref{fig:galaxy}). Each panel is 2\,kpc on each side showing projected image of a 500\,pc-thick slab of the ISM. The symbols and colors are identical to Fig. \ref{fig:c00}. Panel A shows two clouds on the left and right to the center moving toward each other, which merge shortly in panel B. A third cloud above the center is moving toward the central complex and finally merges with it in panel C. In panel D, feedback breaks the complex into two pieces, pushing them apart. A high-surface-density cloud is formed at the center of the panel due to a combination of feedback-driven pressure from the bottom left and cloud collision from the top. The surface density of the cloud is $\sim10^4\,\Msun\,\pc^{-2}$ during cluster formation (panel E). Feedback from the cluster finally blows out the remaining and surrounding gas (panel F).}
\label{fig:c13}
\end{figure*}

\subsection{Cluster Identification}
\label{sec:finder}
When the re-simulations end, we run the cluster finding algorithm\footnote{The code is publicly available at \url{https://github.com/omgspace/Phinder}.} developed by \cite{Grudic:2018a} on all star particles to identify star clusters in the simulated galaxies. The algorithm is summarized as follows. First, we calculate the gravitational potential generated by all star particles. Next, for each particle, we search its 32 nearest neighbors for the particle at the lowest gravitational potential. We move to this particle and do the same search until a local potential minimum is identified. Therefore, every particle is associated with a local potential minimum and all the particles are grouped into particle associations. Lastly, for each group, we re-compute the gravitational potential generated only by the group members and remove those unbound to this group to get a list of bound clusters and their member particles. We also require all bound clusters to have at least 32 particles. The same force softening lengths as in the simulations are used when we calculate the gravitational potential.

In our analysis below, we regard bound clusters with half-mass radius smaller than $\Rhalf=100\,\pc$ as proto-GC candidates. This radius cut is to exclude the galaxies themselves and multiplicities in our analysis and only affects a small number of objects. We only focus on clusters that are newly formed during the starbursts to get a clean comparison between different star formation models (Section \ref{sec:sf}). As our re-simulations only last for $<100\,\Myr$, we ignore cluster destruction in our analysis. We expect this only has a small effect on low-mass and under-resolved clusters that may be disrupted faster than this time-scale. All cluster properties are calculated at the end of the re-simulations. We list the total stellar mass formed in each restart and the fraction of stars ending up in bound clusters in Table \ref{tbl:restart}. We refer to bound clusters, YMCs, and proto-GCs interchangeably in the text below.

\section{The formation of GC candidates}
\label{sec:formation}
\subsection{GC formation in high-pressure clouds}
\label{sec:form}
In this section, we investigate in what conditions and by what physical processes proto-GCs can form. We do so by studying examples of proto-GC candidates formed in the simulation z5m11c\_hr\_G18, where we re-simulate the starburst in galaxy z5m11c\_hr using the star formation model in \citet{Grudic:2018a}. Approximately $1/4$\footnote{For reference, only 4\% of the stars formed prior to the burst still remain in bound clusters by $z=5$, probably due to a combinations of reasons below. First, a lower fraction of stars were formed in bound clusters at early time, as much fewer massive clusters can form in low-mass galaxies (cf. z5m10\_hr in Fig. \ref{fig:mf}). Second, clusters that are marginally resolved by $\lesssim100$ particles will be numerically dissolved over 100\,Myr (cf. discussion in Section \ref{sec:future}). Lastly, cluster disruption is likely efficient because of strong tidal shocks in the progenitor of this galaxy. The oldest cluster found at $z=5$ was formed at $z\sim10$ (over 600\,Myr ago) with metallicity $\rm [Z/H]\sim-2$.} 
of the stars formed in the burst belong to bound clusters with at least 32 particles ($10^{4.5}\,\Msun$) by the end of the re-simulation. Proto-GCs in other simulations are formed in similar ways.

We begin with an overview of the galactic-scale ISM properties during the starburst. In Fig. \ref{fig:galaxy}, we show projected gas images at six epochs from different stages of the starburst. Each image represents a physical length of 16\,kpc along each dimension. Panel A shows a rapid gas infall from nearly all directions prior to the starburst, building up a large amount of gas in the central region of the galaxy (panel B). At epoch C, a patch of the ISM right next to the center has collapsed to sufficiently high densities to trigger star formation. The galaxy maintains an extremely high gas fraction (close to 80\% within the central 5\,kpc) for the first $50\,\Myr$ of the starburst before the SFR decreases dramatically. The gas is highly turbulent with typical velocity dispersion $\sigma_v\sim100\,\km\,\s^{-1}$, corresponding to Mach number ${\cal M}\sim10$--30 for gas of $10^3$--$10^4$\,K. The average gas surface density is over $\Sigma_{\rm gas} \sim 10^2\,\Msun\,\pc^{-2}$, with certain regions in the galaxy reaching $10^3$--$10^4\,\Msun\,\pc^{-2}$.

Panel D illustrates the process where feedback from stars that formed earlier starts to launch galactic winds, which sweep up the gas nearby and create shell-like structures in the ISM. Remarkably, the winds compress the gas to high pressure and (surface) density where star formation takes place rapidly, as shown in panels D and E. This is like the `positive feedback' driven by active galactic nucleus (AGN) outflows \citep[e.g.][though the engine here is massive stars rather than AGNs]{Silk:2013,Bieri:2015} and the `triggered star formation' model\footnote{The `triggered star formation' model is usually discussed in the context of massive star formation within GMCs, but it has also been proposed for cluster formation \citep[e.g.][]{Elmegreen:2002}.} \citep[e.g.][]{Elmegreen:1977,Palous:1994,Elmegreen:2002a,Whitworth:2002,Tan:2000,Tan:2005,Walch:2014}. Finally, feedback from the starburst eventually blows out almost all the gas from the central region and thus star formation is temporarily suppressed in the galaxy (panel F).

Now we study how proto-GCs form in our simulations. Every particle in the simulation has a unique particle ID. By design, each star particle inherits the same ID from its parent gas particle. Thus we are able to trace the particles belonging to a proto-GC identified at the end of the simulation back to the time before they formed. In Fig. \ref{fig:c00}, we illustrate the formation process of a cluster with mass $\Mc=1.4\times10^7\,\Msun$ identified at $z=5$. The formation time ($t=0$) is defined as the epoch when half of the cluster's member particles have formed (i.e. cosmic time $t=1.121$\,Gyr, redshift $z=5.172$ for this cluster). Each column shows a slab ($2\,\kpc \times 2\,\kpc \times 500\,\pc$, the third dimension refers to the direction perpendicular to the page) in the ISM at the given time centered on the median coordinate of all cluster members (and/or their parent gas particles). The color (grey) maps in the first/third (second/fourth) rows represent the projected gas surface density ($\Sigma=\int \rho \dd l$). These images show the gas distribution in the ISM and highlight the dense structures in this region. We checked our results are not sensitive to the thickness of the slab as long as it is larger than the typical size of these dense structures (a few 10\,pc to 100 pc). The vector fields in the first/third rows show the direction of gas motion with respect to the center of mass of all cluster members (and/or their progenitors). The blue points in the second/fourth rows show all star particles in the slab, regardless of their formation time. The red points highlight the cluster members (after they turn into stars). \newpage

This cluster is formed in a high-pressure cloud compressed by feedback from recent star formation nearby, which is an important channel for forming bound clusters in our simulations. This can be seen from panel D of Fig. \ref{fig:galaxy}, where the white dashed box shows a $2\,\kpc\times2\,\kpc$ region centered on this cluster at its formation time. Fig. \ref{fig:c00} presents the formation process of this cluster in more detail. Panels A and B show a gas flow starting to sweep through this region from the upper right to the bottom left 5--10\,Myr before the cluster formed. In the central region of panel B, we find that the flow has entrained a large amount of dense gas, which will merge shortly with the cloud complex at the center of the frame, as shown by the velocity field. As the gas flow keeps sweeping across and compressing the ISM, a dense cloud has formed by $t=-2\,\Myr$ at the center of panel C, where stars that eventually end up in the cluster start to form at the density peaks. During the time when the bulk of cluster stars form (panels C--E), the cloud maintains a super high surface density exceeding $10^4\,\Msun\,\pc^{-2}$. At such high surface density, young stars newly born in the cloud cannot provide sufficient pressure support from feedback to overcome the self-gravity of the cloud, so the cloud converts almost all of its mass into stars in a cloud free-fall time. In the meanwhile, gas from the left side of the frame keeps feeding the cloud, so the cloud can survive for a few Myrs. At $t=5\,\Myr$ (panel E), feedback from the cluster eventually blows out the surrounding gas and terminates cluster formation. These gas flows will likewise drive cluster formation at nearby locations in the same way as described above.


\begin{figure}
\includegraphics[width=\linewidth]{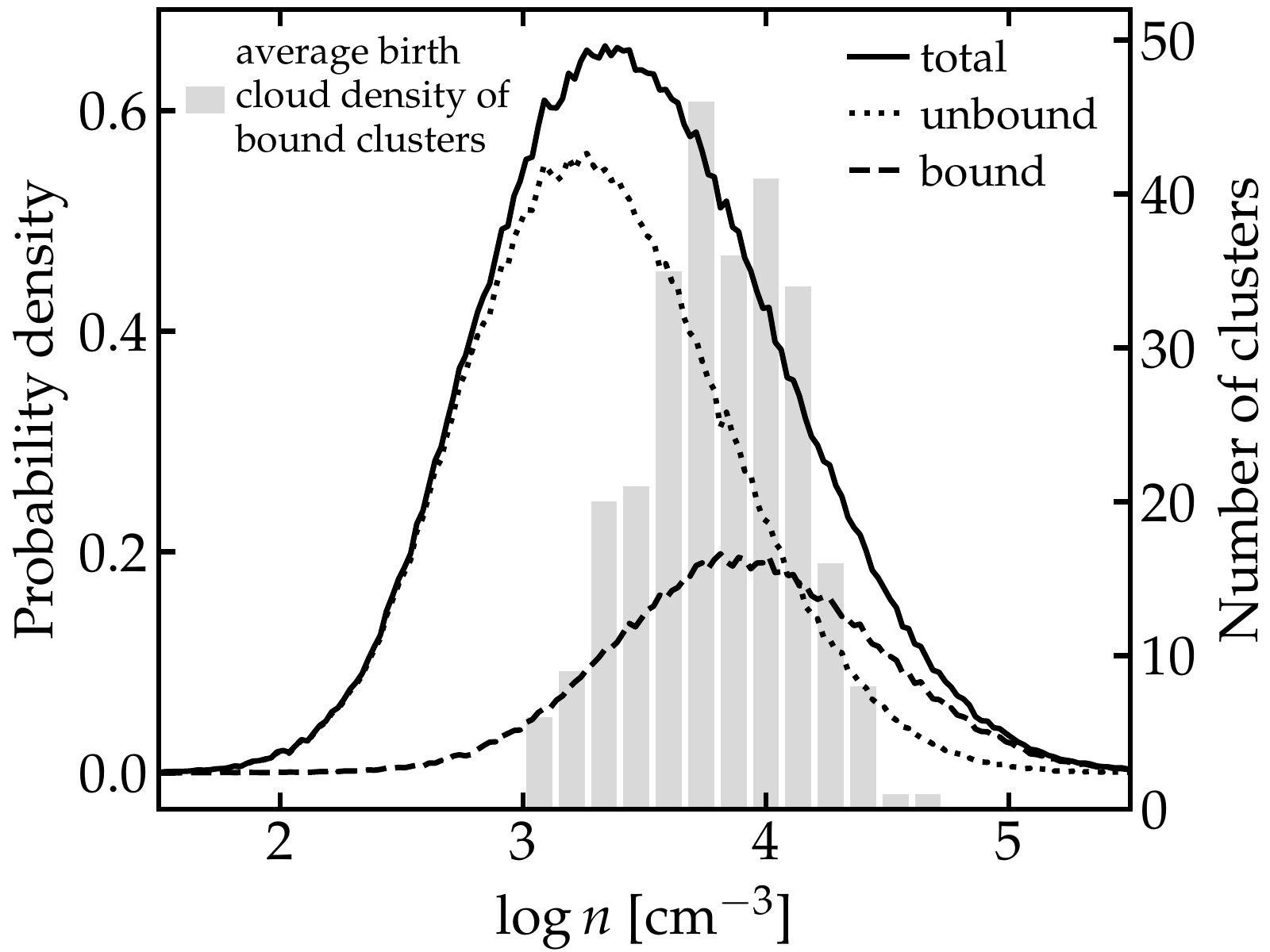}
\caption{Probability distribution function of gas density for the progenitors of star particles formed during the starburst in simulation z5m11c\_hr\_G18, measured in the last snapshot before they turn into stars (solid). The dashed and dotted lines show the contribution from stars belonging to bound clusters and stars not bound to any cluster at the end of the simulation. The grey histogram shows the distribution of average cloud density for all 277 bound clusters formed in the starburst. Cluster stars tend to form in gas an order of magnitude denser than other stars.}
\label{fig:density}
\end{figure}

Fig. \ref{fig:c13} shows the process by which another cluster formed at redshift $z=5.282$ (cosmic time $t=1.092$\,Gyr). The cluster has a final mass $\Mc=1.6\times10^6\,\Msun$ by $z=5$. The white box in panel C of Fig. \ref{fig:galaxy} shows where it forms in the galaxy. Again, each panel represents a slab of $2\,\kpc \times 2\,\kpc \times 500\,\pc$ in the ISM. The symbols and colors are identical to those in Fig. \ref{fig:c00}. We find that at $t=-10\,\Myr$ (panel A), two cloud complexes (left and right of the image center) are moving toward each other and about to collide. A third complex above the center is also falling to the central region. In panel B, the two clouds at the center have already collided and become a larger complex in which star formation is triggered (see the blue points in the bottom panel). In the meanwhile, the cloud above center keeps moving toward the center and eventually merges with the central cloud complex in panel C. In panel D, the complex breaks into two pieces, as feedback acting at the lower left is pushing them apart (see the velocity fields). The piece at the center of frame reaches a high pressure due to a combination of feedback-driven pressure from the bottom left and cloud collision from the top. The cloud maintains a surface density $\sim10^4\,\Msun\,\pc^{-2}$ during the time the majority of cluster stars form (panels D and E). Finally, at $t=6\,\Myr$, the cluster blows out the surrounding gas. High-quality images and animations for Figs. \ref{fig:galaxy}--\ref{fig:c13} are available at this URL\footnote{\url{http://www.tapir.caltech.edu/~xchma/HiZFIRE/globular/}}.


Our simulations suggest a scenario where bound clusters form efficiently in gas-rich, turbulent galaxies with strong star formation activity. The large amount of gas needed can be supplied by rapid gas accretion as commonly happened in the high-redshift universe and/or gas-rich mergers \citep[e.g.][]{Li:2004,Maji:2017}. The turbulence is driven by rapid gas inflows (converting gravitational energy to kinetic energy) and feedback processes. In regions where bound cluster form, external pressure provided by feedback-driven winds and cloud/complex/stream collision must be sufficiently high to compress the gas into high-pressure clouds with surface densities $\gg10^3\,\Msun\,\pc^{-2}$ \citep[see also][]{Elmegreen:2002}. Note that this is much larger than the average gas surface density in the galaxy, indicating that compression by external pressure is important to achieve such conditions. Once the clouds become self-gravitating, they will turn nearly all of their mass into stars in a cloud free-fall time and stars formed in these clouds are gravitationally bound at birth.

\begin{figure}
\centering
\includegraphics[width=\linewidth]{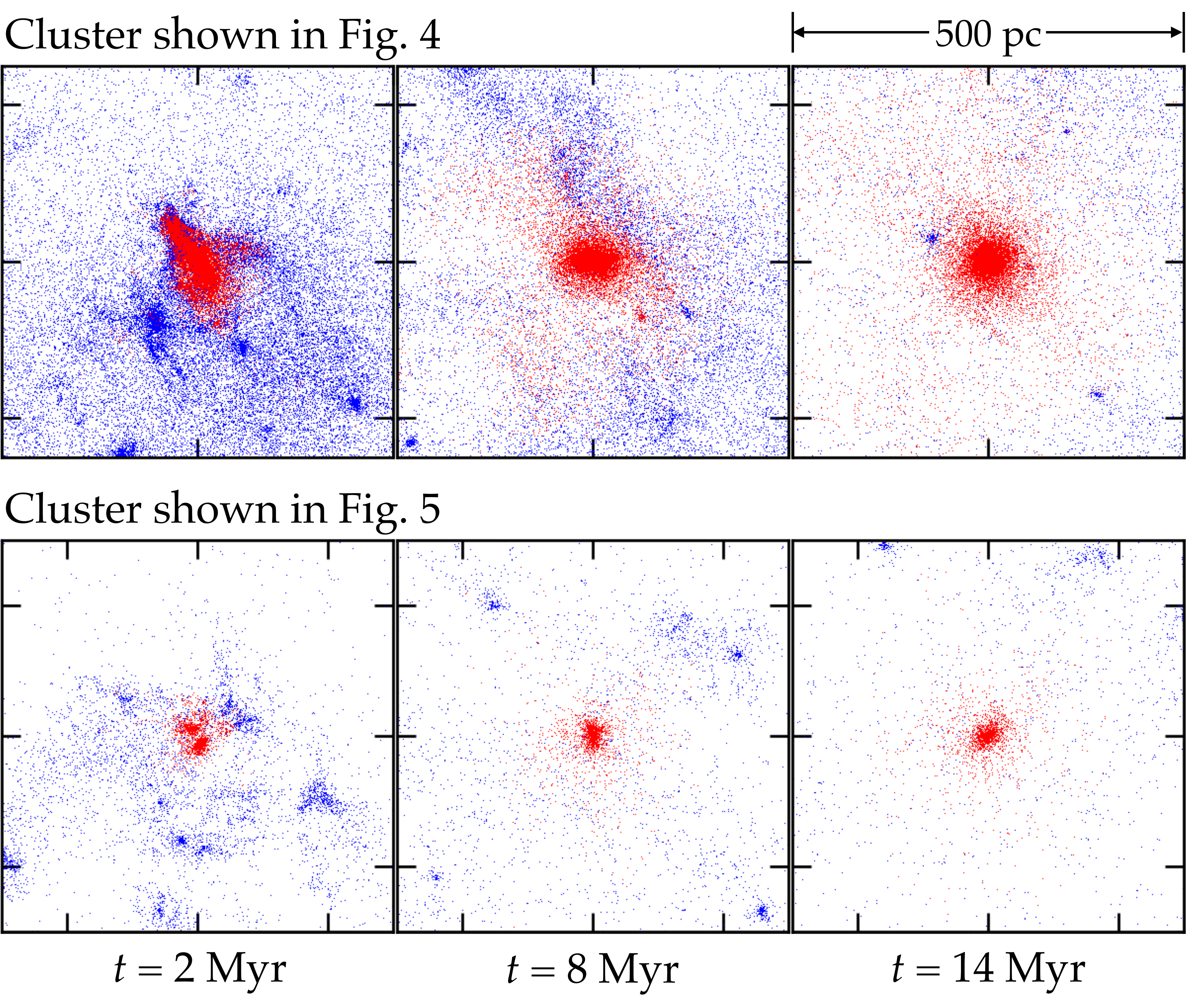}
\caption{Morphological evolution of the two clusters in Figs. \ref{fig:c00} (top) and \ref{fig:c13} (bottom). Each panel shows the projected image of all star particles in a $\rm(500\,pc)^3$ box (blue points) centered on the clusters. Cluster member particles are highlighted by the red points. The clusters are irregular when they form, but quickly relax and become round in shape in about $10\,\Myr$. This process is likely artificially fast because of the modest number of particles in our clusters relative to real systems.}
\label{fig:evolve}
\end{figure}

\begin{figure*}
\includegraphics[width=0.85\linewidth]{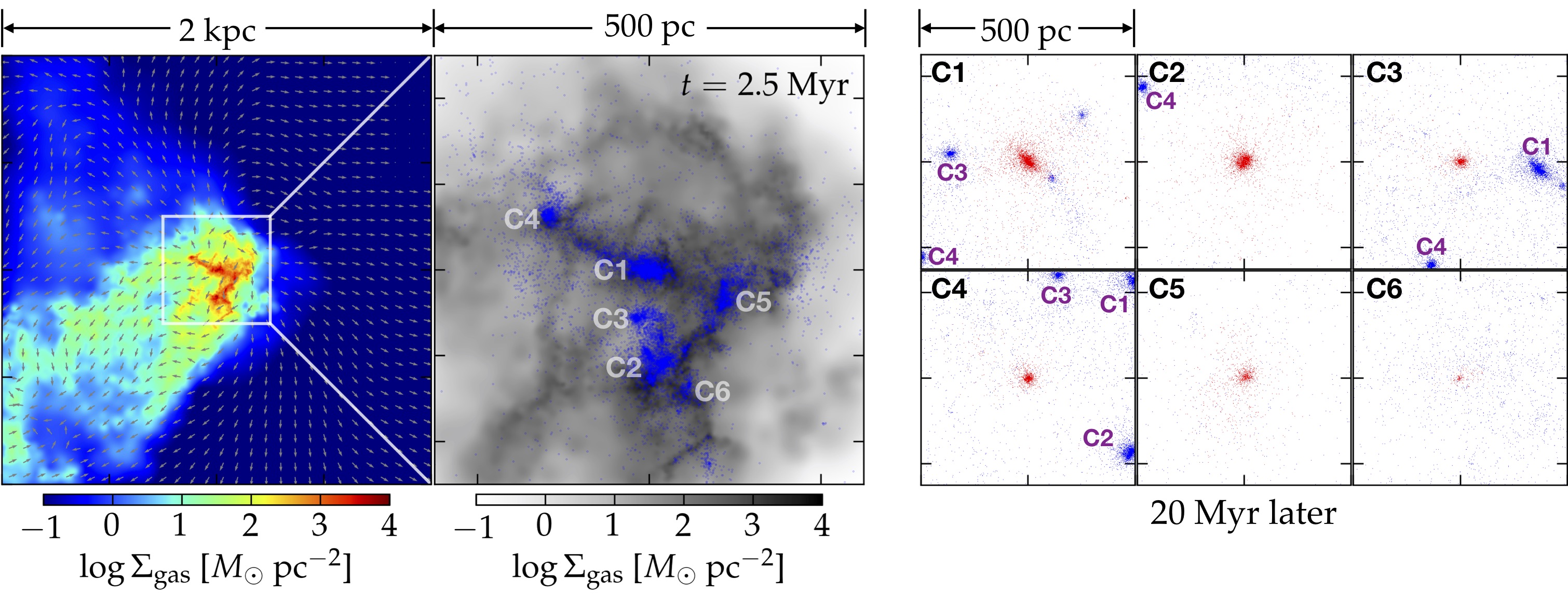}
\caption{An example of `top-down' cluster formation: a cloud complex turns into several bound clusters which disperse from each other. {\em Left}: Projected gas image of a $2\,\kpc\times2\,\kpc\times500\,\pc$ slab in the ISM. A cloud complex is compressed to very high density by a gas flow. {\em Middle}: Projected gas (grey scale) and stellar (blue points) images zoomed into the central $500\,\pc\times500\,\pc$ on the left panel. Six components (labeled by C1--C6) are identified in the newly formed YMC, which become bound clusters by the end of the simulation. Their masses range from $\Mc=3\times10^6$--$1.5\times10^5\,\Msun$ in descending order from C1 to C6. {\em Right}: Clusters C1--C6 at 20\,Myr after their formation. Each panel is $500\,\pc\times500\,\pc$. The blue points show all star particles in the $(500\,\pc)^3$ box, while the red points highlight the members of the cluster of interest. The distances between these clusters increase relative to 20\,Myr ago.}
\label{fig:c06} \vspace{3pt}

\includegraphics[width=0.85\linewidth]{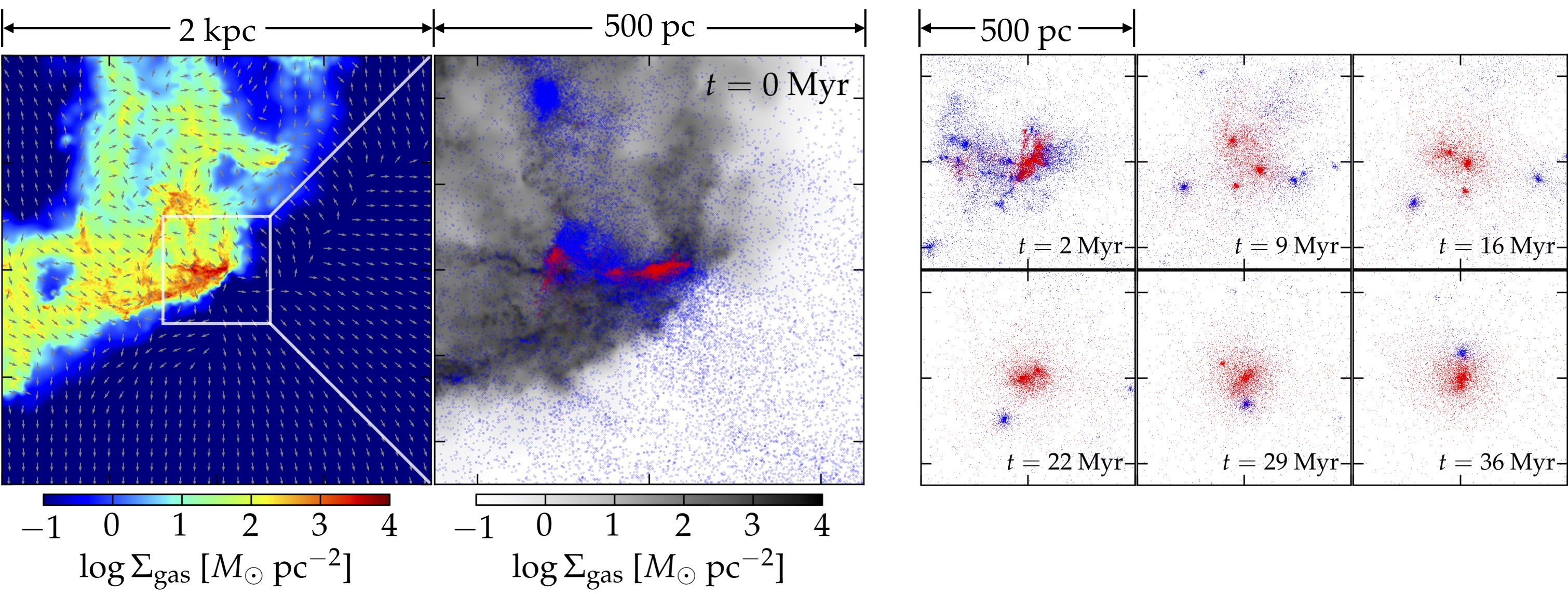}
\caption{A cluster of mass $\Mc=6\times10^6\,\Msun$ formed `bottom-up': a cloud forms several small clusters which merge shortly to form a bigger cluster. The left and middle panels are similar to Fig. \ref{fig:c06}, except that the cluster members are highlighted by red points in the middle panel. The right panel shows the cluster image at six epochs after its formation. Each smaller panel is $500\,\pc\times500\,\pc$. There are three clusters as seen at $t=9$ and 16\,Myr that eventually merge together to form a more massive cluster by $t=36\,\Myr$.}
\label{fig:c03}
\end{figure*}

In Fig. \ref{fig:density}, we further demonstrate that bound clusters preferentially form in high-pressure, high-density regions. For every star particle formed in simulation z5m11c\_hr\_G18, we find the density of its parent gas particle in the final snapshot prior to its birth (the time separation between two successive snapshots is 0.5\,Myr). The solid curve in Fig. \ref{fig:density} shows the probability density function for the `birth density' of all stars formed during the starburst.\footnote{It is highly non-trivial to associate individual star particles with a well-defined `birth cloud' in our simulations, so here we use a local quantity, the birth density of each star particle, rather than cloud-scale properties (which are also non-trivial to define) for diagnostics. We expect the density of star-forming gas in a cloud to correlate with cloud pressure in our simulations, hence stars forming at high densities also means they form in high-pressure (high-surface-density) regions \citep[see also][]{Kruijssen:2012b}.} 
The dashed and dotted curves break the distribution function into stars belonging to a bound cluster and stars not bound to any cluster by the end of the simulation. Stars that end up in bound clusters tend to form in one-order-of-magnitude denser gas than other stars. We checked that the birth density distribution of cluster stars does not strongly depend on cluster mass, at least in the range of $\Mc\sim10^5$--$10^7\,\Msun$. \linebreak For each cluster, we compute the median birth density of all member stars as a substitute for the average density of its birth cloud. In Fig. \ref{fig:density}, the grey histogram shows the distribution of this density for all 277 clusters formed in the re-simulation. This is consistent with the density distribution of all cluster-forming gas (the dashed line). Stars formed at much lower densities (lower-pressure regions) may be unbound in the first place owing to lower cloud-scale star formation efficiencies ($\sim1$--10\% as oppose to near unity in high-pressure clouds) or form marginally bound objects that will be shortly disrupted by feedback. Our findings are in line with previous analytic models that suggest bound clusters form in the most high-pressure, high-density regions in a turbulent ISM \citep[e.g.][]{Elmegreen:1997,Elmegreen:2002,Elmegreen:2012,Kruijssen:2012b}.

The clusters in Figs. \ref{fig:c00} and \ref{fig:c13} do not morphologically resemble GCs in the Universe when they just formed. In Fig. \ref{fig:evolve}, we show the morphological evolution of these two clusters after their formation. The top (bottom) row shows the cluster in Fig. \ref{fig:c00} (\ref{fig:c13}) at $t=2$, 8, and 14\,Myr. Each panel shows the projected image of all star particles (blue points) in a $(500\,\pc)^3$ box centered on the cluster. The cluster members are shown by the red points. Both clusters are irregular at $t=2\,\Myr$, but they quickly become spherical over the next 10\,Myr. We note that this morphological relaxation process happens faster in our simulations than in reality because a cluster is only resolved by a moderate number of particles (e.g. the cluster of mass $\Mc\sim1.5\times10^7\,\Msun$ in Fig. \ref{fig:c00} contains $\sim2\times10^4$ particles). The half-mass radii of the two clusters are $\Rhalf=36.9$ and 23.2\,pc, respectively, by $t=14\,\Myr$. This suggests that clusters formed in high-pressure clouds tend to be gravitationally bound at birth.

\subsubsection{Hierarchical cluster formation}
\label{sec:hierarchical}
In the examples shown in Figs. \ref{fig:c00} and \ref{fig:c13}, almost all stars in the cluster form collectively in a high-pressure gas cloud. Nonetheless, a cloud can form a star cluster complex that may break into several bound clusters at a later time. On the other hand, multiple clusters formed in the same cloud complex can merge together to become a more massive cluster. This is similar to the `top-down-and-bottom-up' hierarchical cluster formation picture discussed in \citet[][and references therein]{Grudic:2018a}. We also identify such processes in our cosmological simulations, which we will explicitly show here.

Fig. \ref{fig:c06} presents an example of multiple clusters forming `top-down' in one cloud. The left panel is 2\,kpc along each side, showing the projected gas image of a 500\,pc-thick slab in the galaxy, similar to the top panels in Figs. \ref{fig:c00} and \ref{fig:c13}. The middle panel shows the image zoomed in on the central $500\,\pc\times500\,\pc$. The grey scale shows the gas surface density and the blue points show all the star particles in the $(500\,\pc)^3$ box. The central cloud complex is compressed by a gas flow in a similar way as in Fig. \ref{fig:c00}. The complex forms a YMC where we identify six components labeled by C1--C6 in the middle panel, which become six bound clusters by $z=5$. We define $t=0$ as the formation time of cluster C1. At this time, all six components are close to each other, within a 150\,pc-radius sphere from C1. The right panel shows the images of the six clusters 20\,Myr later. Each panel is $500\,\pc\times500\,\pc$ centered on the cluster of interest. The blue points show all star particles in the $(500\,\pc)^3$ box, with cluster stars highlighted by the red points. Although clusters C1--C4 are still in close proximity to each other (see e.g. the bottom-left panel), the distances between them increase relative to 20\,Myr ago, indicating that they are falling apart. Clusters C5 and C6 are even further away from C1. The masses of the six clusters range in $\Mc=1.5\times10^5$--$3\times10^6\,\Msun$ (with decreasing mass from C1 to C6).

Fig. \ref{fig:c03} provides an example of `bottom-up' cluster formation, where a bound cluster of mass $\Mc=6\times10^6\,\Msun$ identified at $z=5$ is formed via multiple cluster mergers. The left panel shows the gas image of a $2\,\kpc\times2\,\kpc\times500\,\pc$ slab in the ISM. The middle panel shows the image zoomed in on the central $500\,\pc\times500\,\pc$. The blue points show all star particles in the $(500\,\pc)^3$ box, with cluster stars highlighted by red points. The right panel shows the images of this cluster at six subsequent epochs. Each panel is 500\,pc on each side. There are at least three clusters seen at $t=9$ and 16\,Myr that merge together and form a more massive cluster by $t=36\,\Myr$. Note that the three progenitor clusters are formed in the same cloud complex compressed by a gas flow, so the final cluster is formed `top-down' first and then `bottom-up'.

Although we identify both `top-down' and `bottom-up' cluster formation in our simulations, it is worth noting that these processes are not physically different. Whether a complex forms clusters `top-down' or `bottom-up' is set by the velocity structure of individual cluster-forming clumps in the complex, which determines whether the clusters will fly apart or become bound and merge together. We also stress that `top-down' or `bottom-up' process does not necessarily happen every time a bound cluster forms.

\begin{figure}
\centering
\includegraphics[width=\linewidth]{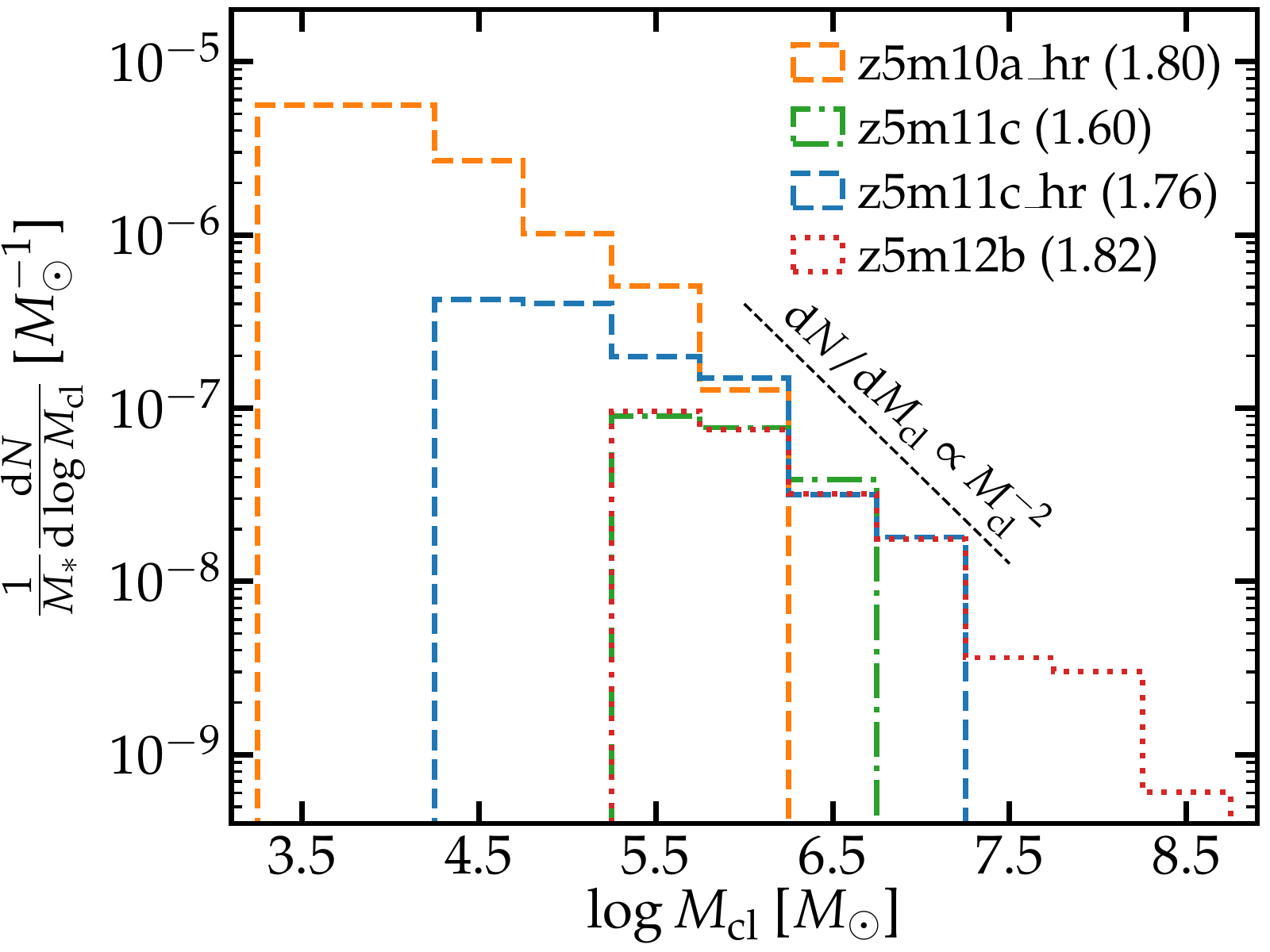}
\caption{Mass function of bound clusters formed in the starburst selected for every galaxy, re-simulated with the default `G18' star formation model. The histograms show the number of clusters per logarithmic mass, normalized by the total stellar mass formed during the starburst. The black dashed line shows the canonical power-law mass function $\dd N/\dd\Mc \propto\Mc^{-\alpha}$ with $\alpha=2$, which our simulations are broadly consistent with. The actual slope $\alpha$ for each run is labeled in the parenthesis. All galaxies have similar efficiencies of forming clusters of a given mass, i.e. forming a cluster of mass $\Mc$ requires roughly a total stellar mass $20\,\Mc$ formed in the galaxy. The good agreement between different simulations suggests the shape of cluster mass function and cluster formation efficiency are robust to resolution.}
\label{fig:mf}
\end{figure}


\subsection{Mass function of newly formed clusters}
\label{sec:mf}
In Fig. \ref{fig:mf}, we present the mass function of bound clusters formed in the starbursts re-simulated with our default `G18' star formation model. For every galaxy, we show the number of clusters per logarithmic mass, $\dd N/\dd\log\Mc$, divided by the {\em total} stellar mass formed during the starburst. To guide the eye, we also show the power-law mass function, $\dd N/\dd\Mc \propto\Mc^{-2}$, with the black dashed line.\footnote{We stress that here we study the mass function of bound clusters newly formed within a short time period. We do not attempt to connect this to the mass function of present-day GCs in this paper. Doing so requires evolving the clusters over cosmic time (including internal dynamics), which is not possible at our current resolution but worth future studies (see Section \ref{sec:future} for more discussion).} This is consistent with the observed YMC mass functions \citep[e.g.][and reference therein]{Zhang:1999,Portegies-Zwart:2010} and in agreement with cosmological simulations \citep[e.g.][]{Kravtsov:2005,Li:2017}. A power-law slope of $-2$ is generically arises from scale-free structure formation \citep[e.g.][]{Guszejnov:2018}. We only consider clusters with at least 32 particles. The mass function is shown every 0.5\,dex in logarithmic mass. We also fit our data by $\dd N/\dd\Mc \propto\Mc^{-\alpha}$ and label the actual slopes $\alpha$ in Fig. \ref{fig:mf}.

In all four galaxies, the newly formed bound clusters follow a mass function broadly consistent with $\dd N/\dd\Mc \propto\Mc^{-2}$. Moreover, we find that all galaxies have similar cluster formation efficiencies, which we refer to the number of bound clusters with a given mass formed per star formation in the galaxy over the same period. This is different from the fraction of stars formed in bound clusters as in Table \ref{tbl:restart}, because the latter does not fully converge with resolution. For example, forming a cluster of mass $\Mc\sim10^{6\pm0.25}\,\Msun$ requires a total stellar mass $\sim2\times10^7\,\Msun$ formed in the galaxy. The stellar mass required to form a cluster of mass $\Mc$ increases linearly with $\Mc$ following the power-law mass function with a slope of $-2$. The variation between different galaxies is within a factor of 2--3. This suggests that our results on the cluster mass function and formation efficiency are not sensitive to the resolution of these simulations, at least when they are run with the same adaptive, resolution-free star formation model. In Section \ref{sec:sf}, we will study how the details of the star formation prescriptions affect our results. 

All the four galaxies are highly gas-rich and turbulent during the starbursts we study here. The bound clusters are formed in similar ways as we present in Section \ref{sec:form}. We caution that the tentative universal cluster formation efficiency found above may only apply to starburst galaxies with similar gas fraction and the extent of turbulent support to those in our simulations. Galaxies in the early universe preferentially meet such conditions, while low-mass galaxies at later times are likely less efficient in forming bound clusters \citep[e.g.][and reference therein]{Elmegreen:2018}. We defer a detailed investigation on cluster formation efficiency over a broad range of galaxy mass, redshift, and ISM properties to a future study.

\begin{figure}
\centering
\includegraphics[width=\linewidth]{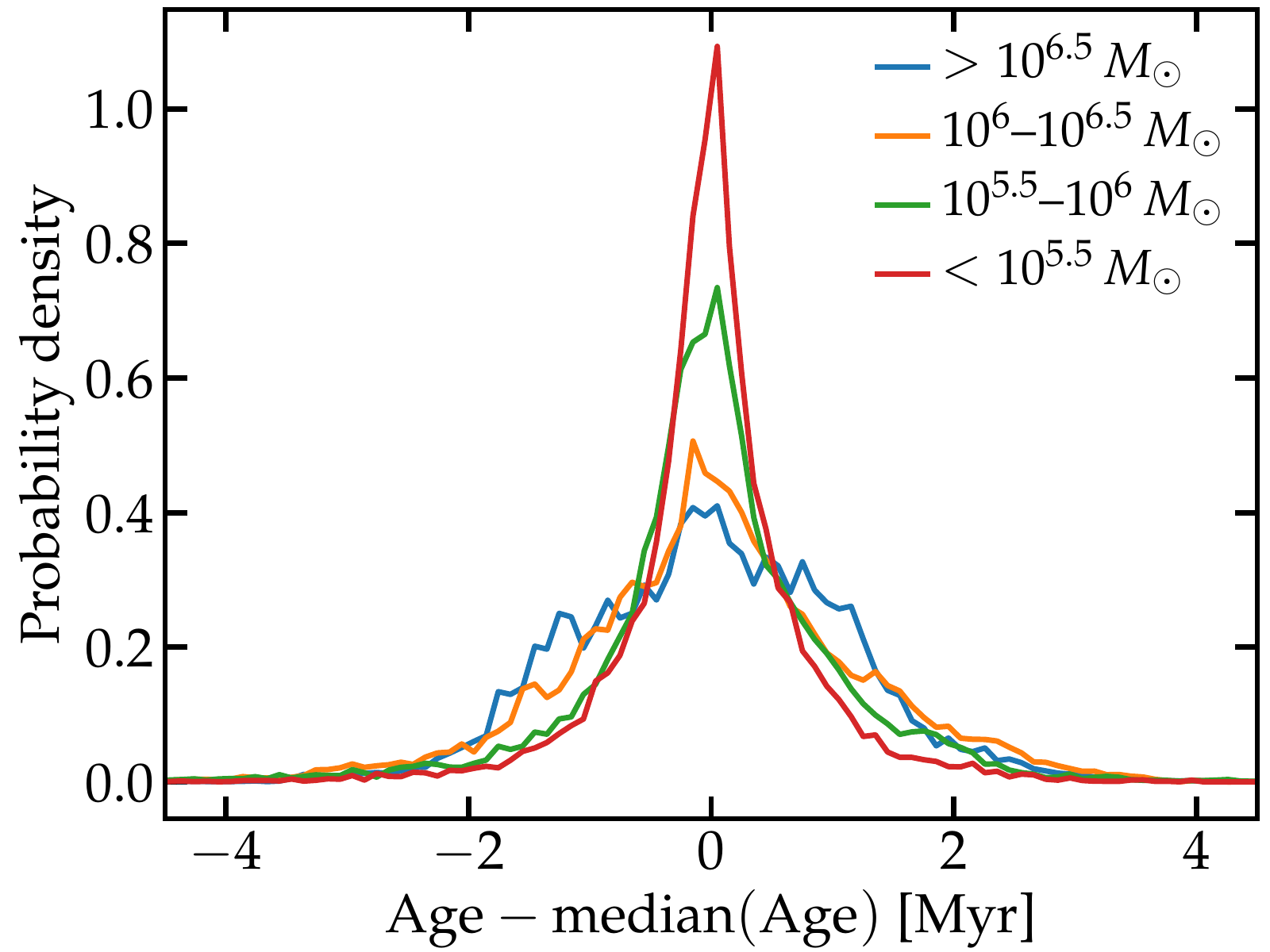}
\caption{The stacked age distribution of cluster members in simulation z5m11c\_hr\_G18, divided in several cluster mass intervals as labeled. The spread of stellar ages increases with cluster mass, with standard deviation increasing from $\sim1\,\Myr$ for clusters below $\Mc\sim10^{5.5}\,\Msun$ to $\sim3\,\Myr$ for clusters above $\Mc\sim10^{6.5}\,\Msun$. This trend holds for all simulations. The typical age spread is a few Myrs in all simulations.}
\label{fig:age}
\end{figure}

\section{Properties of bound clusters}
\label{sec:property}
In Section \ref{sec:form}, we show how bound clusters form in high-pressure, high-surface-density clouds in gas-rich, turbulent ISM. We present the properties of these clusters in this section. We only show results from simulations using the `G18' star formation model. All cluster properties are measured at the end of the re-simulations.

\subsection{Age and metallicity distribution}
\label{sec:age}
Fig. \ref{fig:age} shows the stacked age distribution of cluster stars in bound clusters formed in simulation z5m11c\_hr\_G18 for several cluster mass intervals as labeled. The spread of stellar ages increases with cluster mass from $\sim1\,\Myr$ for the central 68\% stars around the median stellar age in clusters below $\Mc\sim10^{5.5}\,\Msun$ to roughly 3\,Myr in clusters more massive than $\Mc\sim10^{6.5}\,\Msun$. This is broadly consistent with the fact that the spread of ages in nearby YMCs is found to be a few Myrs \citep[e.g.][]{Blum:2001,Mac-Low:2004,Hollyhead:2015} and with the picture where stars form rapidly at near unity efficiency within a free-fall time of the high-pressure clouds \citep[e.g.][and references in Section \ref{sec:intro}]{Grudic:2018}. Note that Fig. \ref{fig:age} essentially shows the SFR of the cluster. The increasing SFR in the early stage of cluster formation means the cluster grows its mass super-linearly, consistent with previous work \citep[e.g.][]{Murray:2015,Murray:2017,Li:2018}.

\begin{figure}
\centering
\includegraphics[width=\linewidth]{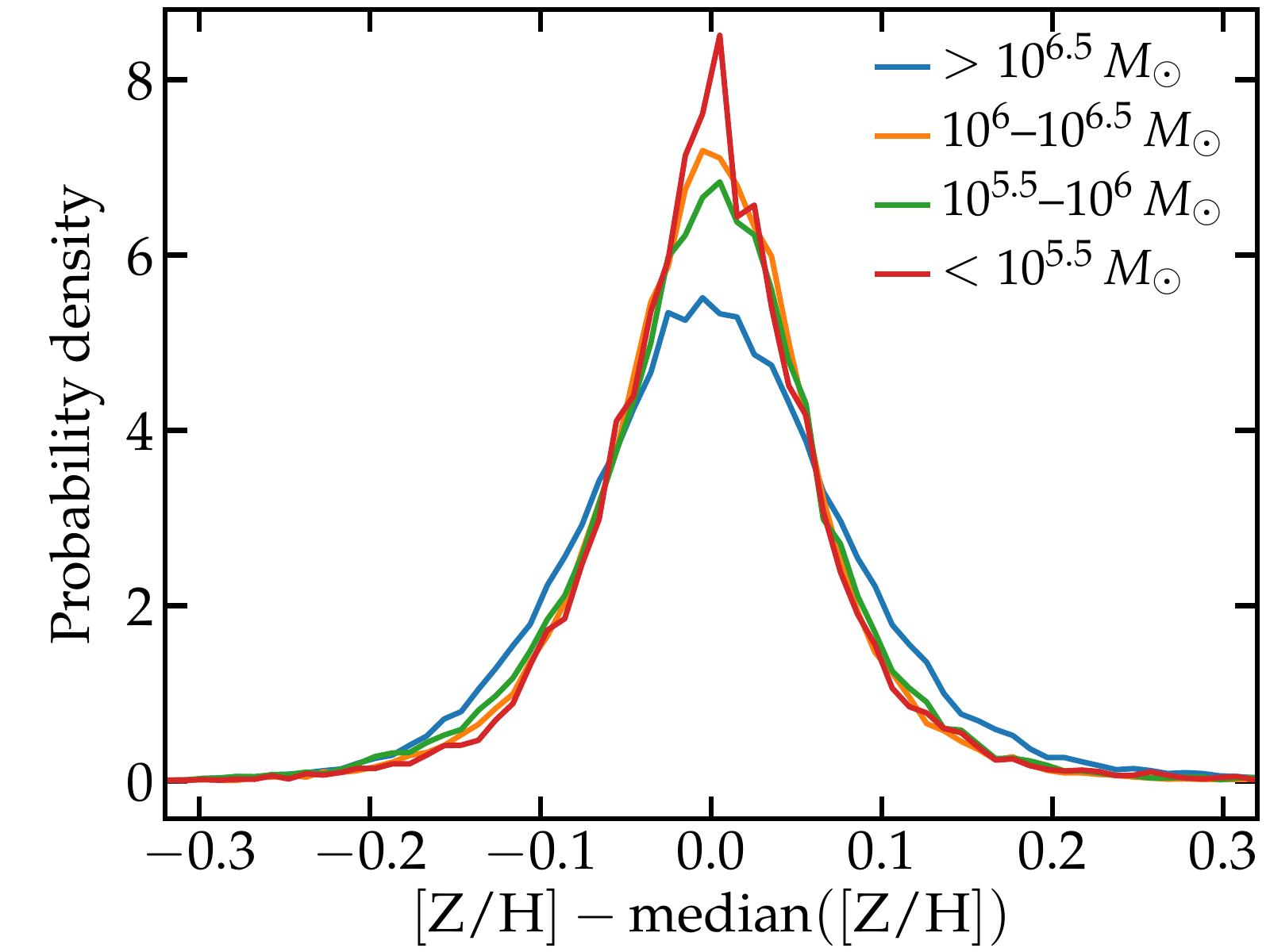}
\caption{Stellar metallicity distribution of cluster members in simulation z5m11c\_hr\_G18, stacked by the same cluster mass bins as Fig. \ref{fig:age}. The $1\sigma$ dispersion of stellar metallicity is $\sim0.08\,\dex$ and does not strongly depend on cluster mass. For individual clusters, the $1\sigma$ dispersion in $\rm [Z/H]$ ranges in 0.03--0.12\,dex. In other simulations, the dispersion of stellar metallicity is also $\sim0.08$\,dex and independent of cluster mass.}
\label{fig:z}
\end{figure}

Fig. \ref{fig:z} shows the stacked metallicity distribution of cluster stars in simulation z5m11c\_hr\_G18 for the same cluster mass bins. The stacked distribution shows a $1\sigma$ dispersion of stellar metallicity in $\rm [Z/H]$ of $\sim0.08\,\dex$ and does not depend on cluster mass significantly. For individual clusters, the $1\sigma$ dispersion in $\rm [Z/H]$ ranges in 0.03--0.12\,dex. This is because most stars in the cluster are born in the same parent cloud with strong turbulence, where metal mixing is expected to be efficient. Our results suggest that cluster stars tend to have near uniform abundance at birth. The observed abundance variations in old GCs may be resulted from stellar evolution effects or subsequent generation of star formation, which still remains an open question \citep[e.g.][for a recent review]{Bastian:2018}. The metallicity of star clusters broadly traces the gas-phase metallicity when and where they formed. We find that the mean metallicity of all bound clusters formed during the starburst in z5m11c\_hr\_G18 has a median $\rm [Z/H]\sim-0.90$ and $1\sigma$ dispersion 0.12\,dex, following the gas-phase mass--metallicity relation at $z=5$ at stellar mass $\Ms\sim10^9\,\Msun$ \citep[e.g.][]{Ma:2016a}. 

We have also checked z5m10a\_hr\_G18 and z5m12b\_G18. We find that in all simulations, the spread in stellar ages increases with cluster mass and is of order a few Myrs. We caution that the trend where the age spread increases with cluster mass is valid only for clusters formed in the same galaxy. We do not compare the age spread for clusters in different simulations, because they are run at different mass resolution. Moreover, we find the metallicity spread is independent of cluster mass in all simulations and the $1\sigma$ dispersion in $\rm [Z/H]$ is $\sim0.08\,\dex$ in all galaxies.

\subsection{The size--mass relation}
\label{sec:size}
\begin{figure}
\centering
\includegraphics[width=\linewidth]{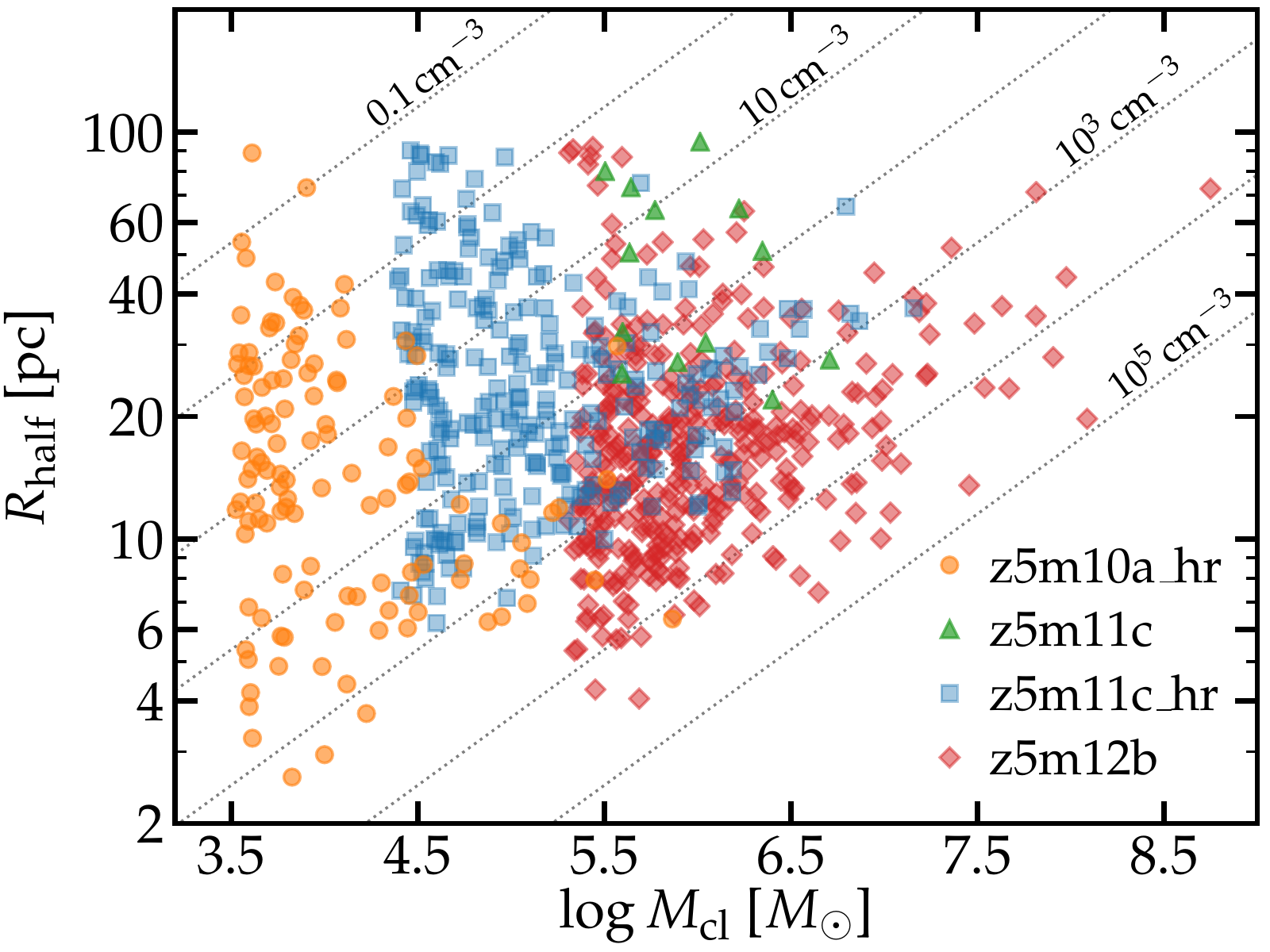}
\caption{The size--mass relation for bound clusters formed in our simulations (in the starbursts re-simulated using the `G18' star formation model). The black dashed lines show constant-density lines from $0.1 m_{\rm H}\,\cm^{-3}$ on the top left to $10^5 m_{\rm H}\,\cm^{-3}$ on the bottom right. Clusters with $\sim100$ particles or less exhibit large scatter in their sizes, probably owing to artificial relaxation for a small number of particles and artificial disruption.} 
\label{fig:size}
\end{figure}

In Fig. \ref{fig:size}, we present the size--mass relation for all bound clusters formed during the starburst we re-simulate for each galaxy with the G18 star formation model, where $\Rhalf$ is the half-mass radius. The black dotted lines represent constant-density lines on this diagram, $\rho=3\Mc/8\pi\Rhalf^3$, from $0.1\,m_{\rm H}\,\cm^{-3}$ on the top left (where $m_{\rm H}$ is the mass of a hydrogen atom) to $10^5\,m_{\rm H}\,\cm^{-3}$ on the bottom right. We only show clusters with more than 32 star particles. Most clusters have a half-mass radius $\sim6$--40\,pc. This is systematically larger than present-day YMCs/GCs and those formed in high-resolution cloud-scale simulations (e.g. Grudi{\'c} et al. 2019). We suspect this is mostly because we have finite mass resolution and force softening lengths in our simulations such that the cuspy density profile in the inner region of the clusters are not well resolved. 

\begin{figure}
\centering
\includegraphics[width=\linewidth]{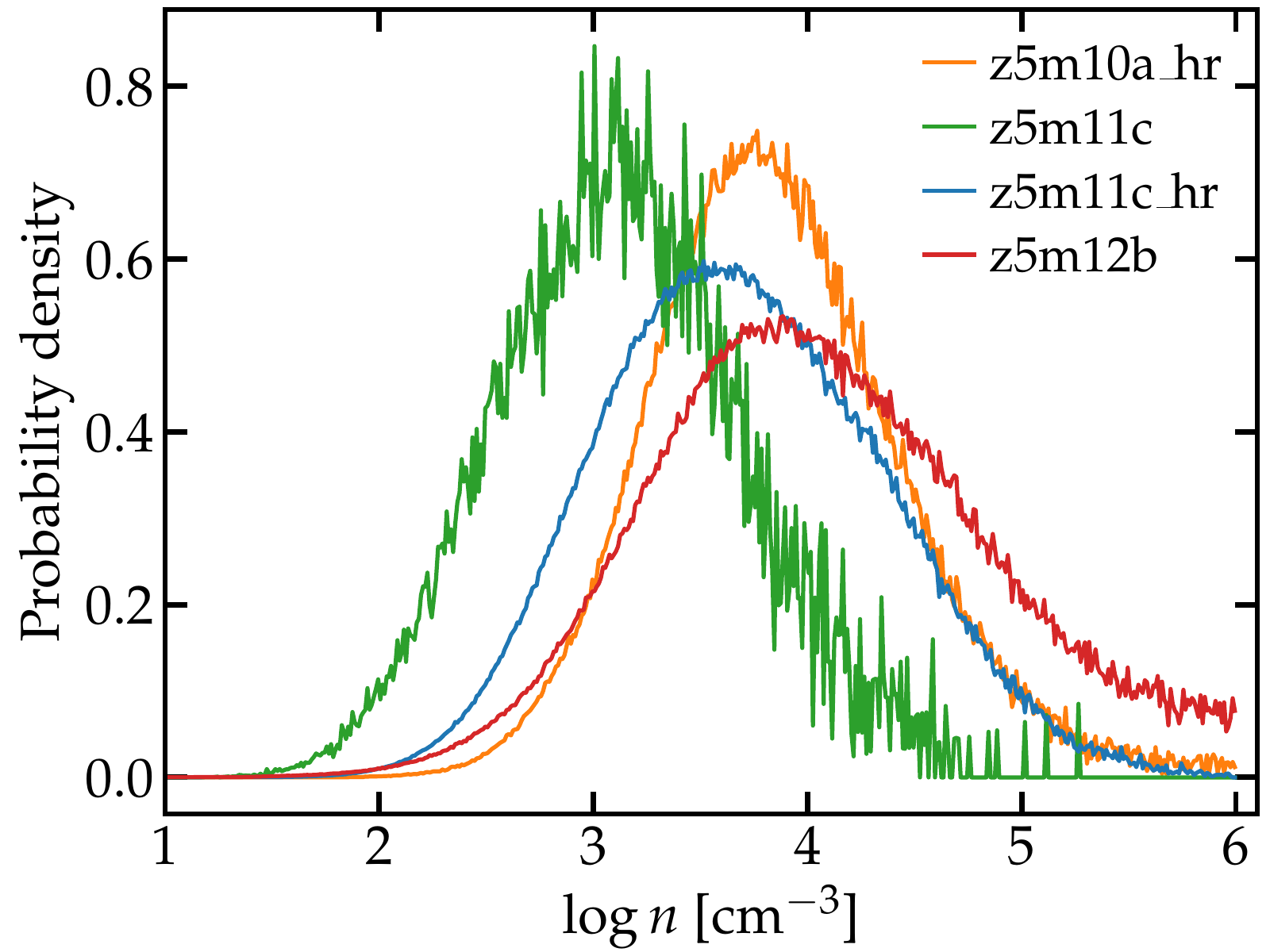}
\caption{The density distribution of star-forming gas (i.e. SFR-weighted distribution of gas densities for all gas particles in all snapshots during the starburst). Star formation tends to take place in denser gas in more massive galaxies (e.g. z5m11c vs. z5m12b at the same resolution) and in simulations run with higher resolution (e.g. z5m11c vs. z5m11c\_hr).}
\label{fig:sfden}
\end{figure}

We note that the density of clusters differs from simulation to simulation, depending on the galaxy mass and mass resolution. For example, clusters in z5m12b have systematically higher densities than those in other simulations, while clusters in z5m11c have the lowest densities on average. To understand this further, in Fig. \ref{fig:sfden}, we show the distribution function of density for all star-forming gas particles (weighted by SFR) in these galaxies.\footnote{Here we take all gas particles within the galaxy from all snapshots during the starbursts we re-simulate and show their SFR-weighted density distribution. This is technically different from Fig. \ref{fig:density}, where we take all star particles formed during the starburst and show the non-weighted density distribution of their progenitor gas particles right before their formation. However, given that star particles are created stochastically from the SFRs of gas particles, the results between Figs. \ref{fig:density} and \ref{fig:sfden} are conceptually identical.} We reiterate that cluster stars tend to form at the high-density end of this distribution (Fig. \ref{fig:density}). We find that the trend in the average cluster density among these simulations broadly follow that in the density of star-forming gas. This can be qualitatively understood as cluster density broadly traces the average density of its birth cloud\footnote{The stellar density of the cluster is a factor of a few lower than the average gas density of its birth cloud, because not all gas in the cloud turns into stars and the stars expand in radius when the remaining gas is expelled by feedback (see e.g. panel F in Figs. \ref{fig:c00}--\ref{fig:c13}).}, which correlates with the density of star-forming gas (where the cloud fragments). At the same resolution (e.g. z5m11c vs. z5m12b), more massive galaxies form stars in denser gas and hence clusters are more compact. This is because z5m12b has a higher average ISM density than the other galaxies. Moreover, better resolution tends to restrict star formation to higher gas densities and the clusters are smaller in sizes (z5m11c vs. z5m11c\_hr). Since we can resolve fragmentation until the local turbulent Jeans mass $M_J\sim \sigma_v^3/(G^{3/2}\rho^{1/2})$ becomes comparable to the mass of a particle $m_b$. Combined with $\sigma_v\sim h^{1/2}$ following energy cascade in supersonic turbulence and $m_b\sim\rho\,h^3$ ($h$ is the resolution scale), this leads to $\rho\sim m_b^{-1/2}$. This is why stars tend to form at a factor of $\sim2\sqrt{2}\approx2.83$ higher densities in z5m11c\_hr than in z5m11c following a factor of 8 difference in mass resolution. We will further discuss the effects of resolution in Section \ref{sec:sf}.

At the low-mass end, where the clusters are only marginally resolved by $\sim100$ particles or less, Fig. \ref{fig:size} shows a large scatter in $\Rhalf$. This is likely because of artificial N-body relaxation owing to a small number of particles and/or these clusters being disrupted too easily (also artificial due to under-resolved densities). In short, we caution that cluster sizes in our simulations depend on the mass resolution, especially for clusters than only contain a small number particles,  whose sizes are heavily affected by numerical effects.

\section{Effects of star formation model}
\label{sec:sf}
In this section, we investigate how the star formation model used in our simulations affects bound cluster formation. We note again the star formation criteria studied here but refer to Section \ref{sec:method} for details: (i) molecular (\mol), (ii) self-gravitating (\sg), based on the local virial parameter, (iii) a density threshold of $\nc=10^3\,\cm^{-3}$ (\den), and (iv) converging flow (\cf). If some or all of these criteria are met, a gas particle is eligible to convert to a star particle at a rate $\dot{\rho}_{\ast}=\eff \, f_{\rm mol} \,\rho/t_{\rm ff}$, where $\eff$ is the {\em local} star formation efficiency of the star-forming particle. Our default choice is $\eff=1$.

We compare four different star formation prescriptions as follows: (1) the FIRE model, which includes criteria {\mol+\sg+\den}, (2) the `no $\nc$' model, which only includes criteria {\mol+\sg}, (3) the `G18' model, which consists of criteria {\mol+\sg+\cf}, and (4) the `G18\_e50' model, which is the same as the `G18' model, except that we adopt $\eff=0.5$ (50\%). Models (1)--(3) adopt the default $\eff=1$. Model (2, `no $\nc$') tends to adaptively pick up overdense structures we are able to resolve for star formation. Model (1, `FIRE') adds a density threshold to model (2), but the effects of the threshold are not resolution independent \citep[see][and discussion below]{Hopkins:2013b}. In practice, a higher $\nc$ is usually applied at ultra-high resolution \citep[e.g.][$\mb\sim30\,\Msun$]{Wheeler:2018}. Model (3, `G18') is also stricter than model (2), but in contrast to model (1), it is also adaptive to resolution. Model (4, `G18\_e50') forms stars at a factor of 2 lower rate at a given density compared to model (3), which means star formation is delayed. We choose model (3) as our default model in Sections \ref{sec:formation} and \ref{sec:property} as it reproduces the fraction of stars formed in bound clusters as observed in a suite of high-resolution star-forming cloud simulations similar to MW and M51 GMCs. It also converges fast with resolution (see Grudi{\'c} et al. 2019, in preparation).

\begin{figure}
\centering
\includegraphics[width=\linewidth]{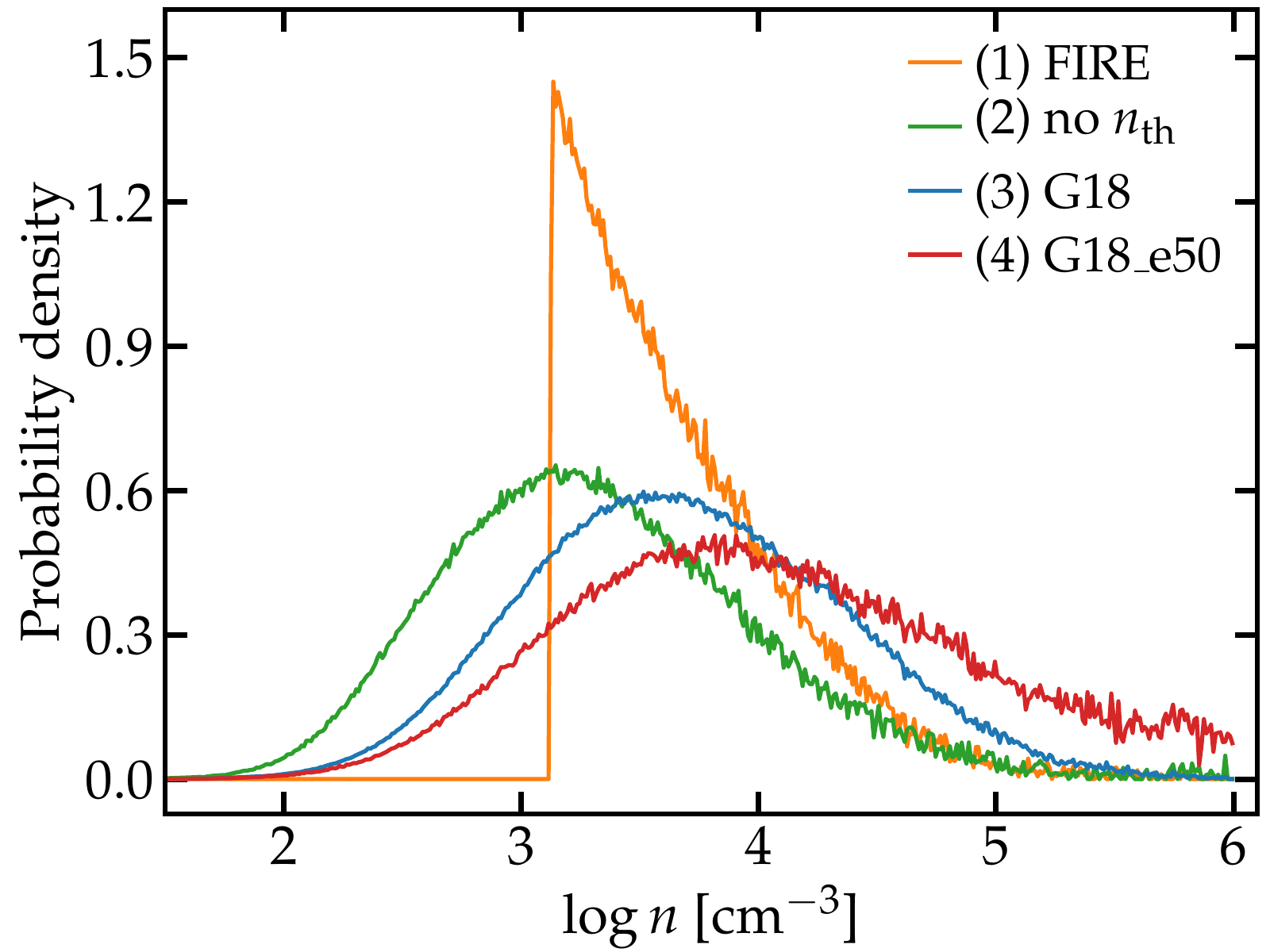}
\caption{Density distribution of star-forming gas from re-simulations of the starburst in galaxy z5m11c\_hr using the four star formation models (see text). Stricter star formation criteria push star formation to higher densities.}
\label{fig:sfcomp}
\end{figure}

The first diagnostic we analyze here is the probability density function of gas density for star-forming gas\footnote{Here we compare simulations of the same starburst in galaxy z5m11c\_hr re-run from identical initial condition, so the difference is entirely due to the star formation model, not resolution or the mean ISM density (see discussion above in Section \ref{sec:size}).} (weighted by the SFR of each gas particle). We re-run the starburst in galaxy z5m11c\_hr using models (1)--(4) and present the results in Fig. \ref{fig:sfcomp}. As we have discussed above, model (2) is the least strict, where stars may form in gas below $100\,\cm^{-3}$ and the peak density for star formation is at $\sim10^3\,\cm^{-3}$. Model (3) is stricter than model (2), which pushes star formation to a factor of 3 higher densities. In model (4), stars form at lower rates while gas keeps collapsing and thus star formation is further restricted to higher densities.\footnote{We have also tested $\eff=0.1$, where star formation occurs at even higher densities. The simulation then becomes too expensive to finish, since there are a large number of gas particles at $n\gg10^5\,\cm^3$ and the timestep is very small for these particles.} Model (1) is also stricter than model (2), since gas below $10^3\,\cm^{-3}$ cannot form stars until it collapses to reach the density threshold. Compared to model (3), model (1) makes it harder for star formation in low-density gas and easier at higher densities.

\begin{figure}
\centering
\includegraphics[width=\linewidth]{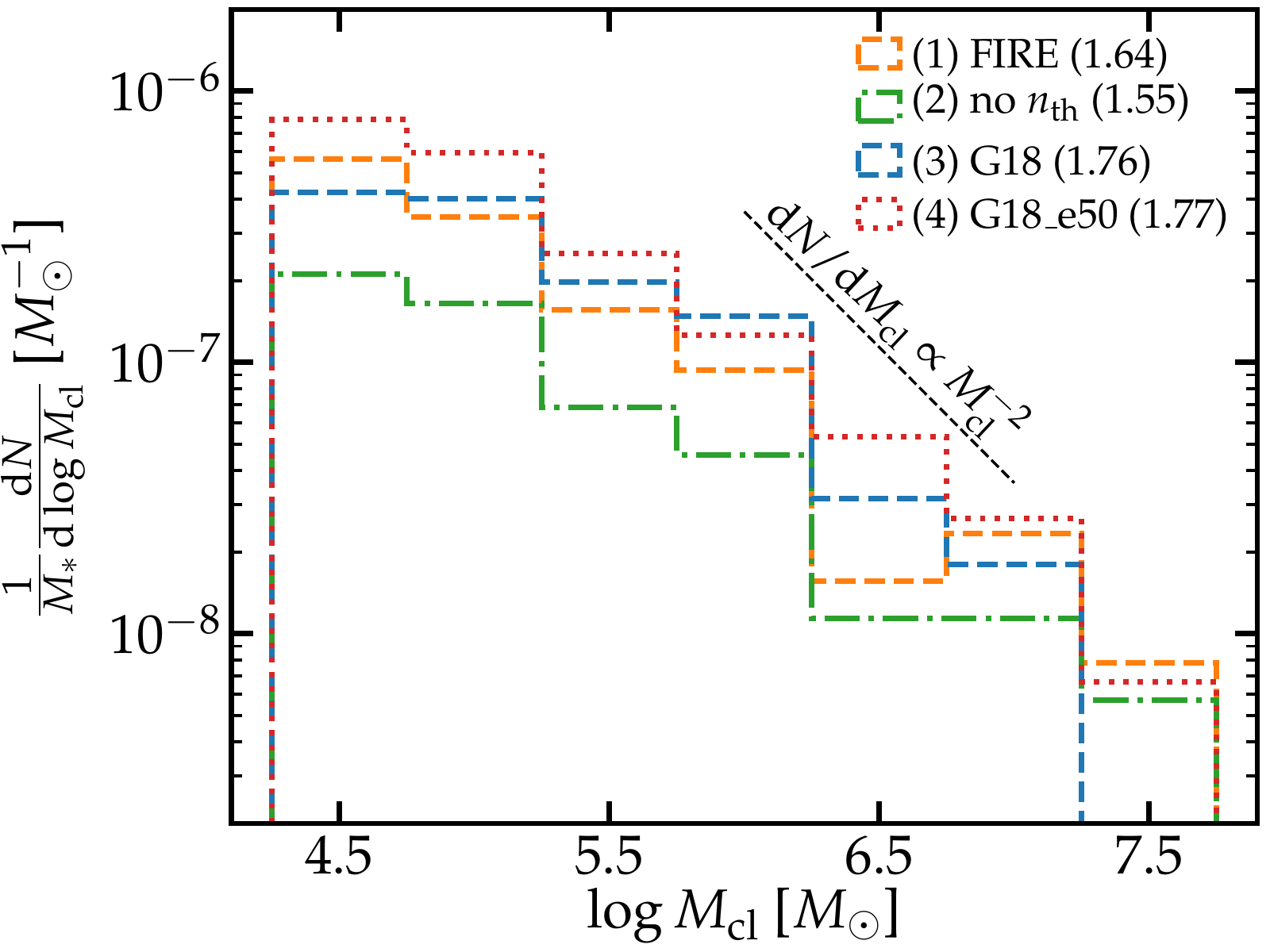}
\caption{Normalized cluster mass function (i.e. number of clusters per stellar mass formed) from the four re-simulations of the starburst in galaxy z5m11c\_hr using different star formation models. The slopes obtained by fitting our data to a power-law function $\dd N/\dd \Mc \propto \Mc^{-\alpha}$ are labeled in the parentheses. As expected from the fact that bound clusters preferentially form in high-density, high-pressure gas, a stricter star formation model results in more clusters per stellar mass formed (i.e. a higher cluster formation efficiency). Nevertheless, we find (a) bound clusters can always form and (b) the cluster mass function broadly follows $\dd N/\dd \Mc \propto \Mc^{-2}$, regardless of star formation models.}
\label{fig:mfcomp}
\end{figure}

In Section \ref{sec:form}, we find that bound clusters preferentially form in high-density, high-pressure gas. Stars formed in low-density gas are likely born unbound in the beginning or in clusters that will be disrupted by feedback shortly. Intuitively, we would expect cluster formation is more (less) efficient when the star formation model is stricter (looser). This is confirmed by the normalized cluster mass functions as shown in Fig. \ref{fig:mfcomp}. Model (2) produces the least clusters at all masses (per stellar mass formed). Models (1) and (3) produce more clusters than model (2) at all masses. Model (4) produces the most clusters at nearly all masses. Compared to model (3), model (1) has more clusters below $\sim10^5\,\Msun$, but less clusters at higher masses. All of these models produce a cluster mass function consistent with $\dd N/\dd \Mc \propto \Mc^{-2}$. We also note again that stochastic effects in the simulations, such as when and where a star particle can form and a SN can occur (stochastically sampled from the rates), can also generate random variations in the (normalized) number of clusters formed in each mass bin. Nevertheless, we find these two findings are robust to our star formation model: (a) bound clusters  form efficiently in high-density, high-pressure clouds in a gas-rich, turbulent galaxy and (b) the newly formed clusters broadly follow a power-law mass function with a slope of $-2$. The fraction of stars formed in bound clusters do, however, depend on the details in the star formation criteria. Models (2), (3), and (4) produce 17\%, 26\%, and 39\% of the stars in bound clusters during the starburst in galaxy z5m11c\_hr (see Table \ref{tbl:restart}). As expected, this fraction increases when the star formation criteria become stricter, as star formation is restricted to higher density gas.

Fig. \ref{fig:mfcomp} shows that a lower $\eff$ actually increases the cluster formation efficiency. This might seem counterintuitive at first glance. We emphasize that $\eff$ here is the {\em local} star formation efficiency in the densest gas following the fragmentation of star-forming clouds down to the smallest resolvable scales in the simulations. Using a lower $\eff$ delays star formation in a cloud, so the cloud can collapse further to higher densities. As the surface densities of star-forming clumps become higher, it effectively {\em enhances} the {\em cloud-scale} star formation efficiencies, resulting in a larger fraction of stars formed in bound clusters. We point out that this trend is opposite to that in \citet{Li:2018}, where the authors find that lowering $\eff$ reduces the cloud-scale star formation efficiencies and thereby the bound fraction. Admittedly, the simulations in \citet{Li:2018} use very different star formation and feedback models from ours and the physical meanings of $\eff$ are likely different between the two studies. Tracking down the exact cause of this discrepancy is beyond the scope of this paper, but it is worth further investigations.

Models (2)--(4) are all adaptive to resolution, while model (1) contains a density criterion which is not necessarily independent of resolution. In Fig. \ref{fig:sfden}, we can see that using the same adaptive star formation model (G18), the density distribution of star-forming gas depends both on galaxy mass (probably the average ISM density if more physical) and mass resolution. Galaxies z5m11c and z5m12b are run at the same resolution ($\mb\sim7000\,\Msun$), but star formation happens in denser gas in z5m12b, likely due to a higher mean ISM density in more massive systems. Galaxies z5m11c and z5m11c\_hr have similar masses (they should be treated as different realizations of the same initial condition) but are run using different resolution ($\mb\sim900$ and $\sim7000\,\Msun$). Star formation in z5m11c\_hr tends to happen in higher-density gas, because we can resolve turbulence on smaller scales than in z5m11c. Therefore, adding a density threshold may have a simulation-by-simulation effect on star/cluster formation. We do not use model (1) as our default in this paper.

\section{Discussion}
\label{sec:discussion}
We have shown that our simulations are able to form bound clusters self-consistently in $z\geq5$ galaxies over a broad range of masses. We show that in a highly gas-rich, turbulent ISM, bound clusters preferentially form in high-pressure regions compressed by feedback-driven winds and collisions between gas clouds/streams. The newly formed clusters broadly follow a power-law mass function with a slope of $-2$ (equal mass per decade in cluster mass). Our results are in line with previous analytic models and numerical experiments, but obtained in cosmological simulations self-consistently for the first time. The clusters formed in our simulations are small in size ($\sim6$--40\,pc in $\Rhalf$, although not as small as real GCs), with cluster stars typically spanning a few Myrs in stellar age and $<0.2$\,dex in metallicity ($\sim0.08$\,dex dispersion in $\rm [Z/H]$), in broad agreement with observed cluster properties. Our findings support the idea that present-day GCs were star clusters/YMCs formed in high-redshift galaxies. We restrict the scope of this paper primarily to how bound clusters form, but briefly discuss some other related topics below. 

\begin{figure}
\centering
\includegraphics[width=\linewidth]{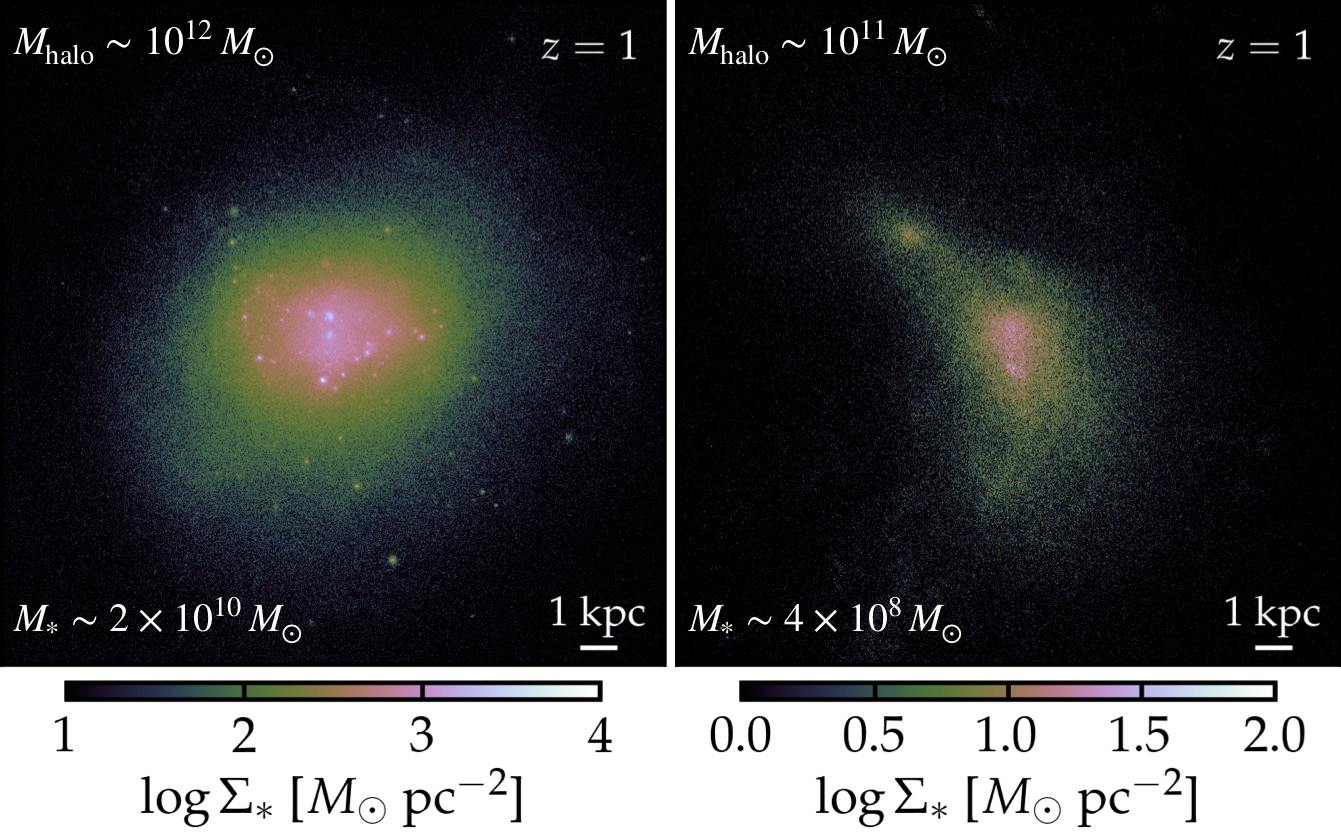} 
\caption{Projected stellar images for galaxy m12i (left) and m11q (right) at $z\sim1$ from the FIRE-2 simulation sample \citep[see][]{Hopkins:2018b}. Their halo mass and stellar mass at this epoch are labeled. Both galaxies have just undergone a starburst. Galaxy m12i can form bound clusters at $z\sim1$. However, no cluster forms in m11q due to lower gas fraction and surface density than high-redshift galaxies at comparable halo mass (e.g. z5m11c\_hr). This suggests bound clusters can form at late times, but they are more difficult to form, especially in low-mass systems.}
\label{fig:m12i}
\end{figure}

\subsection{Cluster formation across cosmic time}
\label{sec:cosmic}
We have focused on a sample of cosmological zoom-in simulations run to $z=5$. However, the key processes we identify for cluster formation are not exclusive to the early universe. As long as the galaxy is gas-rich and turbulent, there is always a large probability to form high-pressure regions due to gas stream collision and/or compression by feedback-driven winds. Such conditions are ubiquitous in high-redshift galaxies, but it may also happen at later times, for example, in gas-rich mergers\footnote{Mergers are also common at high redshifts (e.g. \citealt{Kim:2018} studied a proto-GC formation in a galaxy merger at $z\sim7$), but they are not required to bring in a large amount of gas and stir turbulence. Normal gas infall is sufficient to create a gas-rich, turbulent environment.} and starburst galaxies \citep[e.g.][]{Portegies-Zwart:2010,Oklopcic:2017}. In the left panel of Fig. \ref{fig:m12i}, we show the projected stellar image of simulation m12i from \citet{Wetzel:2016} \citep[also see e.g.][]{Hopkins:2018b} at $z=1$, when the galaxy has just undergone a starburst. The halo mass and stellar mass at this epoch are $\Mhalo\sim10^{12}\,\Msun$ and $\Ms\sim 2\times10^{10}\,\Msun$, respectively, comparable to galaxy z5m12b at $z=5$. This galaxy becomes a MW-mass disk galaxy by $z=0$. This simulation uses the FIRE-2 model for star formation (i.e. model 1 in Section \ref{sec:sf}) and has the same mass resolution as z5m12b. We identify a number of bound clusters formed in this galaxy during the starburst at $z\sim1$ in similar ways as we present in Section \ref{sec:form}.

On the other hand, bound clusters are indeed more difficult to form at lower redshifts. In the right panel of Fig. \ref{fig:m12i}, we present the projected stellar image of another galaxy, m11q, at $z\sim1$, from the FIRE simulation suite \citep[see table 1 in][]{Hopkins:2018b}. Its halo mass at this epoch is $\Mhalo\sim10^{11}\,\Msun$, similar to that of z5m11c\_hr at $z=5$, but it has a factor of 2 lower stellar mass. It is also run with the FIRE star formation model and has the same mass resolution as z5m11c\_hr ($\mb\sim900\,\Msun$). This galaxy has also just undergone a starburst at this epoch. However, we do not find any bound cluster with at least 32 particles formed during the starburst in this galaxy. The peak SFR during the burst is $\sim1\,\Msun\,\yr^{-1}$, almost a factor of 10 lower than that in z5m11c\_hr (cf. Fig. \ref{fig:sfr}). Galaxy m11q also has a much lower gas fraction and gas surface density than z5m11c\_hr at the peak of the starburst.

These results suggest that even at $z\sim1$, cluster formation in low-mass galaxies is less efficient than in the early universe. Below $z\sim1$, it is expected that the gas fraction in galaxies decreases with time and the ISM becomes less turbulent \citep[see e.g.][]{Faucher-Giguere:2013,Hayward:2017}. At the same time, MW-mass galaxies can maintain a stable gas disk \citep[e.g.][]{Ma:2017a,Ma:2017b} in which GMCs are formed mainly via gravitational instability \citep{Toomre:1964}. This is a different mode of star formation from that presented in Section \ref{sec:form}. All of these effects make bound clusters more difficult to form at later times, consistent with the fact that YMCs tend to form in extreme environments at the present day \citep[e.g.][]{Portegies-Zwart:2010}. This is in line with the scenario where present-day GCs preferentially formed at high redshifts, so they tend to be old and metal-poor.

\subsection{How does feedback affect cluster formation?}
\label{sec:feedback}
We do not explicitly study the effects of feedback strength (per stellar mass) on cluster formation in this paper, but briefly discuss them here. It has been suggested that the cluster formation efficiency will decrease when feedback becomes stronger \citep[e.g.][]{Li:2018}. This is plausible since stronger feedback may reduce the star formation efficiency in individual clouds by blowing away more gas and unbinding the stellar association formed in the cloud. 

A higher feedback strength might thus limit bound clusters to form only in the highest-density clouds. On the other hand, it can drive turbulence and generate fast gas flows more efficiently. Such processes may expedite forming high-pressure clouds (via compression) and actually enhance cluster formation in the galaxy. The two effects outlined above are competing with each other and it is not obvious what the net effect is. In our simulations, all feedback quantities (e.g. luminosities, SNe and mass loss rates, etc.) are IMF-averaged and directly computed from stellar population synthesis models. Artificially boosting or decreasing feedback strengths will result in violations of certain observational constraints such as the Kennicutt--Schmidt relation, at least in disk galaxies at $z\sim0$  \citep[see fig. 35 in][]{Hopkins:2018b}. We do not expect the input feedback strength from stars to change by a large factor, but there are certain effects that may enhance the feedback strength per stellar mass, including a top-heavy IMF \citep[e.g.][]{Marks:2012}, stellar rotation and binarity \citep[e.g.][]{Ma:2016}, and radiation pressure due to Lyman-$\alpha$ resonance scattering \citep[e.g.][]{Kimm:2018}. The impact of these on cluster formation is beyond the scope of the current paper but worth studying in a future work.

\subsection{Future directions}
\label{sec:future}
Our high-resolution cosmological zoom-in simulations are able to form bound clusters self-consistently and systematically. However, studying the dynamic evolution of these clusters on cosmological time-scales over 10\,Gyr is a more challenging problem. First of all, at finite mass resolution, clusters in the simulations are resolved by a much smaller number of particles, where $N$-body relaxation happens faster than it should. A cluster may lose a large fraction of its mass or even dissolve on time-scales comparable to 10 relaxation times, which is of order 50\,Myr for a cluster with 100 particles and density $10^2$--$10^3\,m_{\rm H}\,\cm^{-3}$. Secondly, using a finite force softening length, we cannot resolve the central density profile of these clusters, so they can be tidally disrupted more easily in the simulations \citep[see also e.g.][]{van-den-Bosch:2018}. Moreover, as all particles have nearly equal mass in the simulations, some internal dynamic processes due to unequal mass in $N$-body interactions, such as mass segregation, cannot be captured. To reliably trace the dynamic evolution, disruption, and survival of bound clusters over cosmic time in cosmological simulations, it might be necessary to use a hybrid approach where the formation of clusters is explicitly resolved but the dynamics of these clusters are followed by tracer particles \citep[e.g.][]{Li:2017}. Encouraged by previous studies using this method, we expect some clusters formed in the inner region of high-redshift galaxies to be kicked to large radii and survive to $z=0$ \citep[e.g.][]{Kruijssen:2019,Li:2019a}.

We find that the cluster formation efficiency, or the fraction of stars formed in bound clusters of a given mass, in these simulations depends on the star formation criteria. Our default choice (the G18 model) is motivated by high-resolution cloud-scale simulations, but it is not guaranteed that this model is still the `best' application for cosmological simulations. Note that the models studied in Section \ref{sec:sf} do not produce a big difference in global galaxy properties, such as stellar mass, SFR, and star formation history, since they are mainly regulated by feedback \citep[e.g.][]{Hopkins:2013b}. The formation of bound clusters, however, may serve as a test bed for star formation models in cosmological simulations \citep[see also e.g.][]{Li:2018}. 

Last but not least, the results from our simulations are likewise useful for inspiring and calibrating sub-grid cluster formation models for low-resolution cosmological simulations and semi-analytic models \citep[e.g.][]{Choksi:2018,El-Badry:2019}. This is worth studying in the future.

\section{Conclusion}
\label{sec:conclusion}
In this paper, we present the self-consistent formation of bound star clusters in a sample of high-resolution cosmological zoom-in simulations of $z\geq5$ galaxies. For each galaxy, we focus on a starburst which we re-simulate with different star formation models. Our default simulations adopt the FIRE-2 stellar feedback model in \citet{Hopkins:2018b} and a star formation prescription motivated by high-resolution cloud-scale simulations from \citet{Grudic:2018}. We identify the key processes that result in the formation of bound star clusters and study the mass function and basic properties of the bound clusters formed during the starbursts. By comparing simulations of the same starburst with different star formation models, we are able to test the robustness of these results to our star formation model. Our main findings are the following:

(i) We find that bound star clusters preferentially form in high-pressure clouds with gas surface density above $10^4\,\Msun\,\pc^{-2}$, where the cloud-scale star formation efficiencies are nearly unity and the clusters tend to be gravitationally bound at birth (Section \ref{sec:form}, Figs. \ref{fig:c00}--\ref{fig:evolve}). These high-pressure clouds are formed due to compression by feedback-driven winds and collisions of smaller clouds/gas streams in highly gas-rich, turbulent environments (Figs. \ref{fig:galaxy}--\ref{fig:c13}).

(ii) In some cases, clusters may form hierarchically. Multiple clusters may form in one cloud complex and then fall apart (Fig. \ref{fig:c06}, `top-down'). On the other hand, several small clusters born in the same cloud complex may merge to become a more massive cluster (Fig. \ref{fig:c03}, `bottom-up').

(iii) The newly formed bound clusters broadly follow a power-law mass function of $\dd N/\dd \Ms \propto \Mc^{-2}$. In our default simulations, the efficiency of forming a cluster of a given mass is comparable in galaxies in a broad range of mass: one cluster of mass $\Mc$ ($\pm0.25$ dex) forms out of every $20\,\Mc$ of stellar mass formed in the galaxy (Section \ref{sec:mf}, Fig. \ref{fig:mf}). About 25\% of the stars form in bound clusters (Table \ref{tbl:restart}). These results only refer to clusters formed during the starburst we re-simulate for each galaxy.

(iv) The age spread of stars in a cluster is typically a few Myrs and increases with cluster mass (Section \ref{sec:age}, Fig. \ref{fig:age}). The metallicity dispersion of cluster stars is $\sim0.08$\,dex and does not depend strongly on cluster mass (Fig. \ref{fig:z}). The mean metallicity of the clusters broadly traces the gas-phase metallicity of the galaxy when and where the clusters form.

(v) The bound clusters in our simulations typically have half-mass radii in $\sim6$--40\,pc (Section \ref{sec:size}, Fig. \ref{fig:size}), which are considerably larger than real GCs. This means our simulations cannot reliably trace the dynamical evolution of clusters across cosmic time to study present-day GC populations.

(vi) The cluster formation efficiency relies on the star formation criteria used in the simulations. In general, simulations with a stricter (looser) star formation model form more (less) bound clusters per stellar mass formed, though we only find a factor of a few variation for the star formation criteria we consider. A lower local star formation efficiency results in a larger fraction of stars formed in bound clusters (Section \ref{sec:sf}, Fig. \ref{fig:mfcomp}). However, the mass function of newly formed clusters follows a power-law function of slope $-2$ regardless of star formation model.

(vii) Bound clusters can form at lower redshifts, but they are more difficult to form efficiently due to lower gas fraction, turbulent support, and surface density, especially in low-mass systems (Section \ref{sec:cosmic}, Fig. \ref{fig:m12i}). This is in line with the scenario that present-day GCs formed at high redshifts, so they tend to be old and metal-poor.

(viii) The fraction of stars formed in bound clusters and their sizes in our simulations are not fully converged. However, the fact that bound clusters preferentially form in high-pressure regions, the power-law mass function with a slope of $-2$ for newly formed clusters, and the small age and metallicity spread of cluster stars are all robust.

Our findings support the picture that present-day GCs were in fact regular YMCs formed in high-redshift galaxies during normal star formation activity.

\section*{Acknowledgement}
We acknowledge our referee, Oleg Gnedin, for a constructive report that helps us correct a few errors and inaccurate statements and improve the clarity of this paper. We also thank many colleagues, especially Paul Shapiro and Hui Li, for sending us helpful comments and discussions after this paper appeared on arXiv.
The simulations used in this paper were run on XSEDE computational resources (allocations TG-AST120025, TG-AST130039, TG-AST140023, TG-AST140064, and TG-AST190028). 
This work was supported in part by a Simons Investigator Award from the Simons Foundation (EQ) and by NSF grant AST-1715070. EQ thanks the Princeton Astrophysical Sciences department and the theoretical astrophysics group and Moore Distinguished Scholar program at Caltech for their hospitality and support. 
MG and PFH was supported by an Alfred P. Sloan Research Fellowship, NASA ATP Grant NNX14AH35G, and NSF Collaborative Research Grant \#1411920 and CAREER grant \#1455342.
CAFG was supported by NSF through grants AST-1517491, AST-1715216, and CAREER award AST-1652522, by NASA through grant 17-ATP17-0067, by STScI through grant HST-GO-14681.011, and by a Cottrell Scholar Award from the Research Corporation for Science Advancement.
MBK acknowledges support from NSF grant AST-1517226 and CAREER grant AST-1752913 and from NASA grants NNX17AG29G and HST-AR-14282, HST-AR-14554, HST-AR-15006, and HST-GO-14191 from the Space Telescope Science Institute, which is operated by AURA, Inc., under NASA contract NAS5-26555.
AW was supported by NASA, through ATP grant 80NSSC18K1097 and HST grants GO-14734 and AR-15057 from STScI.
DK was supported by NSF grant AST-1715101 and the Cottrell Scholar Award from the Research Corporation for Science Advancement.

\bibliography{/Users/xchma/reference/library}

\label{lastpage}

\end{document}